\documentclass[10pt]{article}
\usepackage[tbtags]{amsmath}
\usepackage{hyperref}
\usepackage{amssymb,amsthm}
\usepackage{amsfonts}
\usepackage{graphicx}
\graphicspath{{figs/}}
\usepackage[left=2.5cm,right=2.2cm,top=0.5cm,bottom=1.3cm,includeheadfoot]{geometry}
\usepackage{indentfirst}
\usepackage{color}
\usepackage{mathtools}
\usepackage{graphicx}
\usepackage{subcaption}
\usepackage{tikz-cd}
\usepackage{enumitem}
\usepackage{braket}

\usepackage[
  backend=biber,
  style=numeric-comp,
  sorting=none,
  doi=true,
  url=false,
  isbn=false
]{biblatex}
\AtEveryBibitem{%
  \clearfield{number}%
}
\renewbibmacro*{in:}{%
  \setunit{\addcomma\space}}
  \DeclareFieldFormat[article,misc]{volume}{\mkbibbold{#1}} 
  \DeclareFieldFormat[article,misc]{eid}{#1} 
\DeclareFieldFormat[article]{pages}{\mkfirstpage{#1}} 

\renewbibmacro*{journal+issuetitle}{%
  \usebibmacro{journal}%
  \setunit*{\addspace}%
  \iffieldundef{series}
    {}
    {\newunit
     \printfield{series}%
     \setunit{\addspace}}%
  \usebibmacro{volume+number+eid}%
  \setunit{\addcomma\space}%
  \printfield{pages}%
  \setunit{\addspace}%
  \usebibmacro{issue+date}%
  \setunit{\addcolon\space}%
  \usebibmacro{issue}%
  \newunit}

\renewbibmacro*{note+pages}{%
  \printfield{note}%
  \newunit}

\definecolor{cites}{rgb}{0.65, 0.20, 0.46}
\definecolor{chapters}{rgb}{0.25, 0.30, 0.56}
\definecolor{header}{rgb}{0.0, 0.5, 0.64}

\numberwithin{equation}{section}

\hypersetup{
	colorlinks=true,
	linkcolor=header,
	citecolor=cites,
	urlcolor=chapters
}

\def\m0{\bar{M}}

\usepackage{float}
\usetikzlibrary{decorations.pathmorphing}
\usepackage[normalem]{ulem}

\def\D{\mathrm{d}}
\def\I{\mathrm{i}}
\def\Ob{\mathcal{O}}

\def\eq#1{\begin{equation}\begin{aligned}#1\end{aligned}\end{equation}}
\def\eref#1{(\ref{eq:#1})}

\renewcommand*{\thefootnote}{\fnsymbol{footnote}}

\begin{document}
\begin{center}
{\Large\bf Radial quantization of the Schwarzschild geometry:\\ relational observables, evaporation, and remnant state}
		\vskip 5mm
		{\large
			David Brizuela${}^{1,}$\footnote{Contact author: {\tt david.brizuela@ehu.eus}},
			Leonardo Chataignier${}^{2,}$\footnote{Contact author: {\tt lchataig@cbpf.br}},
            and Mikel López-Escondrillas${}^{1,}$\footnote{Contact author: {\tt mikel.lopeze@ehu.eus}}
		}
		\vskip 3mm
		{\sl ${}^{1}$Department of Physics and EHU Quantum Center, University of the Basque Country,\\
			Barrio Sarriena s/n, 48940 Leioa, Spain}\\
			{\sl ${}^{2}$ Centro Brasileiro de Pesquisas Físicas, Rua Dr. Xavier Sigaud 150, Urca, CEP: 22290-180,\\ Rio de Janeiro, RJ, Brazil}\\
\end{center}

\setcounter{footnote}{0}
\renewcommand*{\thefootnote}{\arabic{footnote}}

\begin{abstract}
We consider the canonical quantization of the Schwarzschild geometry in a minisuperspace description based on a foliation with leaves labeled by the radial variable. This formulation remains regular across the Killing horizon. In this way, the resulting quantum theory admits a physical Hilbert space of states
that solve the radial constraint equation and simultaneously describe the exterior and interior regions of the black hole.
We discuss the construction of quantum relational observables, which act as gauge-invariant symmetric operators on the physical Hilbert space.
In order to analyze quantum effects on the geometry, we construct a family of semiclassical intelligent states
that saturate the Robertson--Schr\"odinger uncertainty relation for both the black-hole mass and
the asymptotic Killing norm, while exhibiting small relative fluctuations.
The expectation values of relational observables in these states admit a clear geometric interpretation, and they lead to an effective, quantum-corrected spacetime.
In particular, we find that the position of the horizon may be shifted, depending on the correlation of the state.
In addition, both the proper acceleration of a static observer and the horizon surface gravity are enhanced
with respect to a classical Schwarzschild black hole with the same horizon area.
Using the relation between surface gravity and temperature, we then obtain corrections to the thermodynamic
properties of the black hole, leading to a logarithmically corrected Bekenstein--Hawking entropy.
Finally, we discuss the definition of a state of minimal energy that could be interpreted as a
stable remnant characterized by a finite horizon radius, suggesting a possible mechanism
for preventing the complete evaporation of the black hole.
\end{abstract}

\section{Introduction}\label{intro}

The quantization of black-hole spacetimes is particularly interesting because it may indicate what the full theory of quantum gravity implies for the formation,
evolution, and evaporation of black holes, leading to a richer and more involved toy theory in comparison to the quantization of homogeneous models of cosmology \cite{KieferBook}. In particular, the canonical quantization of spherically symmetric black-hole models allows for a concise, analytical treatment, and it has a long history. 

Kucha\v{r} \cite{Kuchar:1994zk}, as well as Kastrup and Thiemann \cite{Kastrup:1993br,ThiemannKastrup:1993}, performed the canonical quantization of spherically symmetric spacetimes using metric and Ashtekar variables, respectively. In both approaches, the theory was reduced to a finite-dimensional model by explicitly separating the physical degrees of freedom (the Schwarzschild mass and its canonically conjugate quantity) from the gauge degrees of freedom. This was achieved by a series of canonical transformations before quantization. On the other hand, the formal quantization of spherically symmetric spacetimes was carried out without first performing such a separation of degrees of freedom in, for example,
Refs.~\cite{Brotz:97,Kenmoku:1999,Kenmoku:1999bf}. In these approaches, the quantization was rather implemented on a system with an infinite number of degrees of freedom, and formal solutions to the constraints were obtained. Although such a field-theoretic quantization is important, as it is closer to the full quantization of canonical gravity, it is of limited use due to its formal nature. In particular,
in order to have an unambiguous definition of quantum probabilities,
one must still regularize the constraint equations \cite{Christodoulakis:1986,Christodoulakis:1987,Tsamis:1987,Friedman:1988,Christodoulakis:1991}
and find solutions that can be rigorously shown to be normalizable in an appropriate inner product.

The issue of the inner product, along with the general construction of the Hilbert space,
and the definition of observables is much better understood in homogeneous toy models of cosmology,
the so-called minisuperspace models \cite{KieferBook,DeWitt:1967,Misner:1969,Misner:1972,Blyth:1975,Halliwell:91,Ashtekar:94,Landsman:95,Marolf:97,Embacher:98,Giulini:99,Giulini:99-2,Giulini:2000,Marolf:2000,Halliwell,Chataignier:2019kof,Chataignier:2020fys,Chataig:Thesis}. Interestingly, one can describe static, spherically symmetric vacuum spacetimes in terms of a minisuperspace model. As shown in Refs.~\cite{Cavaglia:1994yc, Cavaglia:1995bb}, the radial coordinate can play the role of the Hamiltonian ``evolution'' parameter in these models, unlike in the standard Arnowitt--Deser--Misner (ADM) formulation, in which the evolution parameter is a time coordinate associated with a spacelike foliation.
In Refs.~\cite{Cavaglia:1994yc, Cavaglia:1995bb}, the Wheeler–DeWitt quantization of the system was analyzed,
with special attention devoted to the properties of operators generating rigid symmetries of the Hamiltonian. 

In the present article, we use a canonical formalism in the radial variable, similar to the one presented in \cite{Cavaglia:1994yc, Cavaglia:1995bb} (see also \cite{Christodoulakis:Cavaglia,Melas:2013hoa,Lenzi}),
together with the framework developed in \cite{Chataignier:2019kof,Chataignier:2020fys,Chataig:Thesis} to systematically construct quantum relational observables associated with the Schwarzschild geometry. In contrast to some previous analyses (such as  \cite{Fragolino:PAW,Mann:PAW}), we do not need to assume that a particularly simple type of ``clock'' (or of ``rod'') is present in the model via the so-called Page--Wotters mechanism \cite{PAW}. Rather, the construction used here allows for a more general ``rod'', as long as the associated Faddeev--Popov operator can be suitably defined \cite{Faddeev1,Faddeev2}. Our analysis thus complements the previous works that dealt with a radial minisuperspace treatment of the quantized Schwarzschild geometry (see also \cite{Magueijo,Gielen:2025ovv,Lenzi,Chiba}) by developing the discussion of quantum corrections to the geometry via a systematic treatment of relational observables.
In addition, this canonical formulation provides a smooth evolution across the
Killing horizon. In this way, the physical quantum states we will obtain simultaneously describe 
the exterior and interior regions of the black hole. Therefore, we will analyze in detail
physical effects produced by the quantum degrees of freedom at the horizon and in its surroundings.

The main body of the paper is organized as follows. In Sec.~\ref{sec:ClassTh}, we present the classical theory, including its Hamiltonian formulation, relational observables, classical solutions, and
the characterization of the trapped and antitrapped regions.
Section~\ref{sec:QuanTh} is devoted to the canonical quantization of the model,
paying special attention to the construction of the physical Hilbert space
and gauge-invariant observables.
In Sec.~\ref{sec:states}, we construct specific physical states, such as the
eigenstates of the mass operator, as well as a family of so-called intelligent
states that saturate the Robertson–Schrödinger uncertainty relation.
Using these intelligent states, in Sec.~\ref{sec.physicsintelligentstates},
we provide a semiclassical description of the Schwarzschild black hole,
and discuss its main gravitational and thermodynamic properties.
Finally, in Sec.~\ref{sec:Conclusions} we present a summary of our main results.

\section{Classical theory}\label{sec:ClassTh}

\subsection{The model}
We begin our analysis by considering a spherically symmetric line element,
\begin{equation}\label{eq:line-el}
 \D s^2=g_{ab} \D y^a \D y^b+r^2 \D\Omega^2,
\end{equation}
where $g_{ab}$ is a two-dimensional Lorentzian metric such that Latin indices take the values $0$ and $1$, with $(y^0,y^1)=(t,x)$ and $\D\Omega^{2}=\D\theta^2+\sin^2\theta\, \D\varphi^2$ being the metric of the unit two-sphere. The areal radius function $r$ determines the area of the two-spheres as $4\pi r^2$, and we consider it to be
positive definite $r> 0$. We furthermore assume the existence of a Killing field $\partial_t$, and we parameterize the spacetime metric as
\eq{\label{eq:le}
    \D s^2=
    -\frac{u(x)}{r(x)}\D t^2+2s(x)\D t\D x+q(x)\D x^2+r^2(x)\D\Omega^2,
}
where $u$, $r$, $q$, and $s$ are real functions of the coordinate $x$. The assumption of a Lorentzian signature implies that
the determinant of the metric should be negative, that is,
\begin{equation}\label{eq:lorentziancondition}
   N^2:=-{\rm det}(g)=\frac{u}{r}q+s^2>0 \,.
\end{equation}
As long as this condition is fulfilled, the signs of the functions
$u$, $q$, and $s$ remain unrestricted.

The squared norm of the Killing field $\partial_t$ is given by
\begin{equation}\label{eq.killing}
\xi:=-\frac{u}{r},
\end{equation}
and the causal character of this vector defines the symmetry properties of the corresponding
region. Regions where the vector field is timelike $(\xi<0)$ are static, while regions where it is spacelike $(\xi>0)$ are homogeneous. In addition,
the vanishing of $\xi$ defines Killing horizons, where $\partial_t$ is null.

The metric function $q$, on the other hand, provides the squared norm of the coordinate
vector $\partial_x$. Therefore, the causal character of $\partial_x$ in a given region
depends on the coordinate chart.
However, even if $\partial_x$ may be spacelike,
our goal is to construct a Hamiltonian formulation using the coordinate
$x$ as the ``evolution'' parameter, so $\partial_x$ is to be understood as the generator of the evolution.
This is the specific sense of what we will refer to as ``radial evolution''.
In this way, we will be able to dynamically describe this static spacetime
within a minisuperspace approach. In particular, this
description will be valid across the horizon. This will be of special
relevance in the quantum theory, since
we will be able to consider both the interior and exterior
regions of the black hole simultaneously in a unique Hilbert space,
and derive quantum properties of the horizon.

To this end, we consider the Einstein--Hilbert action over a volume $\cal V$,
\begin{equation}\label{eq:EH}
 S=\frac{1}{16\pi G}\int_{\bf \cal V} \D^4x\,N\,r^2\,\sin\theta\,R,
\end{equation}
where we set the vacuum speed of light $c = 1$, $G$ is Newton's constant, and,
without loss of generality, we choose $N>0$, such that $\sqrt{-{\rm det}(g)}=N$. Taking into account the fact that the Ricci scalar $R$ only depends on $x$, we can integrate the $(\theta,\varphi)$ coordinates over the whole solid angle, while the $t$ coordinate can be integrated over an interval $[t_0,t_1]$. The result is the action
\begin{equation}\label{eq:action-M0}
 S=\frac{M_0}{2} \int_{I} \D x \, N\,r^2\, R\,,
\end{equation}
where $I$ is an appropriate integration interval for $x$ and $M_0:=(t_1-t_0)/(2 G)>0$ defines an energy scale, which appears as a global prefactor in the action. Finally, we explicitly compute the Ricci scalar of the line element in Eq.~\eref{le} and, after integrating by parts, the boundary terms are exactly cancelled by the inclusion of the Gibbons--Hawking--York term, yielding the minisuperspace action
\begin{equation}
    S
    =M_0\int_I \D x\,  \left(\frac{u'r'}{N}+N\right),
\end{equation}
where the prime denotes a derivative with respect to $x$. Note that, although the metric in Eq.~\eref{le} depends explicitly on the functions $q$ and $s$,  their derivatives do not appear in the action. Moreover, these quantities only enter the action in the combination $N$
\eqref{eq:lorentziancondition}. As a result, it will be sufficient to consider $N$ as the relevant configuration variable instead of $q$ and $s$ separately.

\subsection{Hamiltonian formulation}

Taking $x$ as the evolution parameter, the canonical momenta conjugate to $u$ and $r$
are defined in the standard way as derivatives
of the Lagrangian with respect to $u'$ and $r'$, respectively, and they read
\begin{equation}\label{eq:momenta}
\begin{aligned}
p_u &= {M_0}\frac{r'}{N},\\
p_r &= {M_0}\frac{u'}{N}\,.
\end{aligned}
\end{equation}
Since the action does not depend on $N'$, the canonical momentum conjugate to $N$ vanishes, $p_N = 0$. This constitutes a primary constraint, and it reflects the reparameterization invariance inherent to general relativity. As $p_N = 0$, the variable $N$ can be seen as an auxiliary field. The resulting nontrivial phase space is therefore spanned by the variables $u$ and $r$, together with their conjugate momenta $p_u$ and $p_r$. The only nonvanishing Poisson brackets are $\{u,p_u\}=\{r,p_r\}=1$.

Solving the definitions~\eref{momenta} for the velocities $u'$ and $r'$, the canonical Hamiltonian follows from a Legendre transformation and takes the form
\begin{equation}\label{eq:ham}
H=NM_0\left(\frac{p_u p_r}{{M_0^2}}-1\right)\,.
\end{equation}
The requirement that the primary constraint $p_N = 0$ be preserved under the evolution in $x$ generates a secondary constraint,
\begin{equation}\label{eq:constraint}
C:=\left(\frac{p_u p_r}{{M_0^2}}-1\right) = 0\,.
\end{equation}
The auxiliary field $N$ thus serves as a Lagrange multiplier that enforces the secondary constraint $C = 0$. In the standard Rosenfeld--Bergmann--Dirac analysis, no further generation of constraints appear. Consequently, we obtain a first-class (indeed, Abelian) algebra of constraints generated by $p_N$ and $C$. The Hamiltonian \eqref{eq:ham} vanishes on the constraint surface and the dynamics is entirely governed by the constraint $C$.
More precisely, the corresponding equations of motion read
\begin{equation}\label{eq:eom}
\begin{aligned}
 u'&=\frac{N}{{M_0}}\,p_r\,,\\
 r'&=\frac{N}{{M_0}}\,p_u\,,\\
 p_u'&=0\,,\\
 p_r'&=0\,,\\
 C &= 0\,,
\end{aligned}
\end{equation}
with $N$ being an arbitrary but nonvanishing smooth function of $x$. The freedom to choose $N$ corresponds to the gauge freedom of the model.

\subsection{Initial-value problem and Dirac observables}

The gauge invariance of the model is given by transformations of $N$ and the basic phase-space variables that do not change the functional form of the equations of motion~\eref{eom}. Consider the transformations
\eq{\label{eq:gauge}
\delta_\zeta N &= \zeta' \,,\\
\delta_\zeta f &= \{f,\zeta C\}\,,
}
where $\zeta=\zeta(x)$ is a gauge parameter and $f$ is a generic phase-space function (it depends on
$u$, $r$ and their conjugate momenta). It is straightforward to verify that, under $N\to N+\delta_\zeta N$, $f\to f+\delta_\zeta f$, the equations of motion retain their functional form, and thus the transformations \eref{gauge} are a local symmetry of the model.

If we consider a smooth function $\zeta(x)$ that is zero near an initial point $x = x_0$ but nontrivial elsewhere, the local-symmetry transformations \eref{gauge} imply that there exist different solutions to Eqs.~\eref{eom}, $f_{(1)}$ and $f_{(2)} = f_{(1)}+\delta_\zeta f_{(1)}$, which have the same initial data at $x = x_0$, since $\delta_\zeta f_{(1)}|_{x = x_0} = 0$ and $f_{(2)}|_{x = x_0} = f_{(1)}|_{x = x_0}$. In this way, the initial-value problem is not well posed for general phase-space functions, unless they are gauge invariant, as in such case $\delta_\zeta f = 0$ regardless of the form of $\zeta$.

The observables of this model are the quantities that can be unambiguously determined from initial conditions, i.e.,
for which the initial-value problem is well posed. Thus, observables $\Ob$ must be gauge-invariant phase-space functions,
i.e., they should satisfy
\eq{
\delta_\zeta \Ob =0,
}
for every admissible choice of $\zeta$, when the constraint $C = 0$ is enforced. It is clear that this is achieved if $\Ob$ Poisson-commutes with the constraint,
\eq{\label{eq:invariant}
\{\Ob,C\} = 0\,.
}
Quantities that satisfy Eq.~\eref{invariant} are also called Dirac observables and, if Eq.~\eref{invariant} is fulfilled only
on-shell (that is, when the constraint $C = 0$ is enforced), we say that $\Ob$ is a weak Dirac observable. As (weak) Dirac observables have a well-posed initial-value problem, they encode the physical content of the model.
We will discuss below how the classical solutions can be understood in terms of Dirac observables.

In particular, we note that, since for the present (completely constrained) system
the Hamiltonian is proportional to the constraint $H=N M_0 C$, weak Dirac observables
are constants of motion on-shell (when evaluated on solutions), that is, $\dot{\cal O}=\{{\cal O}, H\}=0$
on the phase-space surface $C=0$.

\subsection{\label{sec:cl-sol}Classical solutions}

Let us now explicitly show that the Schwarzschild geometry is recovered as the solution to the dynamical equations \eref{constraint} and \eref{eom}. For such a purpose, it is useful to define the Misner--Sharp mass,
\begin{equation}\label{eq:mf-gen}
 M:=\frac{r}{2G}\left(1-(\nabla_a r) (\nabla^a r)\right)\,,
\end{equation}
where $\nabla$ is the covariant derivative of the 2-dimensional metric $g_{ab}$ defined in Eq.~\eref{line-el}. For the parameterization in Eq.~\eref{le}, the Misner--Sharp mass takes the form
\begin{equation}\label{eq:mf}
    M=\frac{1}{2 {G}}\left(r-\frac{up_u^2}{{M_0^2}}\right)\,.
\end{equation}
It is easy to check that this is a weak Dirac observable,
\begin{equation}
 \{M,C\} = -\frac{1}{2 G M_0^2} p_u C = 0\,,
\end{equation}
where, for the second equality, the constraint $C=0$ is enforced.
As commented above, Dirac observables take
a constant value on solutions.
However, we will indistinctly use $M$ to indicate the phase-space function \eref{mf},
as well as its on-shell value. The difference will be clear from the context.
In fact, unless otherwise stated, from this point on all expressions
are to be understood evaluated on shell.

It is straightforward to integrate the equations of motion~\eref{eom}.
Both the momenta $p_u$ and $p_r$ are seen to be Dirac observables and,
taking into account the constraint \eref{constraint},
they can be written as
\begin{equation}\label{eq:classicalp}
\begin{aligned}
p_u&=M_0 k,\\ p_r&=M_0/k,
\end{aligned}
\end{equation}
where $k\neq 0$ is a dimensionless integration constant.
In fact, $k$ can be regarded as a Dirac observable defined by $k^2:=p_u/p_r$.
Then, these two observables, $k$ and $M$ as defined in Eq.~\eref{mf}, can
be used to solve $u$ as a function of $r$,
\begin{equation}\label{eq:classicalu}
u= (r-2 G M)/k^2.
\end{equation}
At this point, the only remaining equation is $r'=Nk$. One could solve this explicitly by fixing a gauge, that is, choosing a particular functional form for the Lagrange multiplier $N(x)$, which is otherwise undetermined. Alternatively, one can simply express the other fields in terms of $r(x)$. In this relational description, the solutions obtained above relative to $r(x)$, and parameterized by the observables $k$ and $M$,
can be substituted in the metric~\eref{le}.
More precisely, using the definition of $N(x)$ in Eq.~\eref{lorentziancondition} to remove the variable $q(x)$, we obtain
\begin{equation}\label{eq:pre-Schwarzschild}
 \D s^2=-\frac{1}{k^2}\left(1-\frac{2GM}{r}\right) \D t^2+ 2 s \D t \D x+\frac{N^2-s^2}{1-\frac{2GM}{r}}k^2 \D x^2 +r^2 \D\Omega^2\,.
\end{equation}
The Dirac observable $k$, as well as the fields $s\equiv s(x)$ and $N\equiv N(x)$, can be removed by a change of spacetime coordinates. Indeed, by defining a new variable $T$ from
\begin{equation}
\D T=\frac{\D t}{k} -\frac{s(x) k}{1-\frac{2GM}{r(x)}}\D x\,,
\end{equation}
and, by noting that the equation of motion $r' = Nk$ implies $Nk \,\D x = \D r$, the line element \eref{pre-Schwarzschild} takes the standard form in Schwarzschild coordinates,
\begin{equation}\label{eq:Schwarzschild}
 \D s^2=-\left(1-\frac{2GM}{r}\right) \D T^2+\frac{1}{1-\frac{2GM}{r}}\D r^2 +r^2 \D\Omega^2.
\end{equation}
The geometry is thus completely determined by the value of the observable $M$. However,
at the dynamical level, one needs two observables, $M$ and $k$, to parameterize the solutions.
It is only when one reconstructs the geometry that $k$ can be removed. Therefore, as will explained
below in more detail, the observable $k$ contains information about the coordinate choices, and, in
this sense, is observer dependent. However, at the quantum level, we will not have a straight geometric
interpretation at hand, and both Dirac observables $M$ and $k$ will appear as parameters of the quantum
states. In fact $k$ will be used to provide a coordinate representation of the physical states on the
Hilbert space.

\subsection{Relational observables}
\label{sec:clrelob}
A useful class of Dirac observables is relational, in the sense that it refers to the values of quantities relative to a reference field. This is, in fact, what was done in obtaining the line element \eref{pre-Schwarzschild}, where we expressed the solutions to Eqs.~\eref{eom} relative to the $r(x)$ field.

The relational Dirac observables can be constructed as follows. Take a generic phase-space function
$\chi(x)$, and assume that it is monotonic on an open interval $\mathcal{I}\subset I$ of values of $x$. Then, its level sets can be used to define a new coordinate $\sigma$ via
\eq{\label{eq:gf}
\chi(x) = \sigma \,,
}
where, by definition, $\sigma$ takes all the
values $\chi(x)$ for $x\in {\cal I}$. Therefore,
$\sigma$ defines a new parameterization
of the open interval $\cal I$.
In particular,
the monotonicity of $\chi(x)$ in $\cal I$ means that $\chi'(x)\neq0$ in $\cal I$, and we find the resolution of the identity
\eq{\label{eq:FPid}
1 = \int_{\cal I}\D x\,|\chi'(x)|\delta(\chi(x)-\sigma) \,.
}
This is simply the Faddeev--Popov gauge-fixing formula \cite{Faddeev1,Faddeev2}, where Eq.~\eref{gf} functions as a gauge-fixing condition, and the factor of $|\chi'(x)|$ is the analogue of the Faddeev--Popov determinant.
Due to the properties of the Dirac delta function,
we can move this factor out of the integral in
Eq.~\eref{FPid}, and write
\eq{\label{eq:FPid2}
\frac{1}{\Delta_\chi(\sigma)}:=\left.\frac{1}{|\chi'(x)|}\right|_{\chi(x)=\sigma}=\int_{\cal I}\D x\, \delta(\chi(x)-\sigma) \,.
}
This is an invariant quantity because
\eq{\label{eq:FPinv}
\left\{\frac{1}{\Delta_\chi(\sigma)},C\right\}=\frac{1}{N} \int_{\cal I}\D x\, \frac{\D}{\D x}\delta(\chi(x)-\sigma) = 0 \,,
}
where we have used the on-shell identity
$\{\delta(\chi(x)-\sigma),H\}=\frac{\rm d}{{\rm d}x}\delta(\chi(x)-\sigma)$
and the assumption that
$\sigma$ does not take the values
corresponding to the boundaries of ${\cal I}$. From this, general relational Dirac observables can be constructed from the expression
\eq{\label{eq:relobs}
\Ob[f|\chi = \sigma] := \Delta_\chi(\sigma)\int_{\cal I}\D x\, \delta(\chi(x)-\sigma) f(x) \,,
}
which can be shown to be invariant in a manner similar to Eq.~\eref{FPinv}. These invariants encode the value of a phase-space function $f(x)$ when $\chi(x) = \sigma$, thus being relational. We can also rewrite Eq.~\eref{relobs} as
\eq{
\Ob[f|\chi = \sigma] = \int_{\cal I}\D x\,|\chi'(x)|\delta(\chi(x)-\sigma)f(x) = f(\chi^{-1}(\sigma)) \,,
}
where $\chi^{-1}$ is the inverse function associated with $\chi(x)$. So, in particular, we have the trivial relational observable
\eq{
\Ob[\chi|\chi = \sigma] = \chi(\chi^{-1}(\sigma)) = \sigma \,,
}
which expresses the value of $\chi(x)$ relative to itself.\footnote{It answers the tautological question: ``What is the value of $\chi(x)$ when $\chi(x) = \sigma$?''}

For the system under consideration, taking the solutions to Eqs.~\eref{eom} presented in Sec.~\ref{sec:cl-sol}, we see that
\eq{
\Ob[u|r = \sigma] = u(x)|_{r(x) = \sigma, C = 0} = (\sigma-2GM)/k^2 \,,
}
which is the relational solution we used to obtain the line element \eref{pre-Schwarzschild}, only there we did not write $\sigma$
but we simply used $r$ to denote a given value of the radial variable.

In this case,
the choice of $r(x)$ as the reference field is intuitive: it corresponds to choosing the radius as the ``evolution'' parameter relative to which we describe the solutions to Eqs.~\eref{eom}. However, in the quantum theory developed next, it will be advantageous to instead consider $\chi\equiv\ln (r/\ell)$ as the reference field, where $\ell$ is an arbitrary reference length.
Since, from this point on, we will impose $\chi\equiv\ln (r/\ell)$, we
introduce the short-hand notation ${\cal O}_f$ for the observable
corresponding to the phase-space function $f(x)$, that is,
\begin{equation}
{\cal O}_f:={\cal O}[f|\ln(r/\ell)=\sigma]
=f(\ell e^\sigma).
\end{equation}
In particular, the observable corresponding to the areal
radius $r$ now reads
\begin{equation}\label{rel.radius}
\Ob_r =\ell e^\sigma\,,
\end{equation}
and the relational observables associated to the remaining phase-space variables $u$, $p_u$, and $p_r$ are
\eq{
\Ob_u = \frac{1}{k^2}\left(\ell e^\sigma-2GM\right),\,\, \Ob_{p_u}=M_0k,\,\, \Ob_{p_r}=\frac{M_0}{k},
}
which can be obtained by simply replacing $r$ by $\ell e^\sigma$
in the classical solutions \eqref{eq:classicalp}--\eqref{eq:classicalu}. Finally, we see that the relational observable associated with the Misner-Sharp mass is itself, $\Ob_M = M$.

\subsection{Trapped and antitrapped regions}\label{sec.trapped}

Note that we are not imposing a specific sign for $M$, and thus
the metric \eqref{eq:pre-Schwarzschild} describes the Schwarzschild black hole
(for $M>0$), Minkowski (for $M=0$), and the
Schwarzschild naked singularity (for $M<0$).
Unless otherwise stated, all the computations will be valid
for all possible values of $M$. However, we note that only
the case with $M>0$ presents a horizon, and thus, whenever
we mention the horizon we will be assuming a positive $M$.

In particular,
in addition to its form in the Schwarzschild chart \eqref{eq:Schwarzschild},
the line element \eref{pre-Schwarzschild} is rather general and, for $M>0$, it reproduces different horizon-penetrating coordinates, such as Eddington-Finkelstein ($N^2=k^2=1$ and $s=\pm 1$ for either ingoing or outgoing coordinates) and Gullstrand–Painlevé ($N^2=k^2=1$ and $s=\sqrt{2GM/r}$). From this general line element, the squared norm of the Killing field $\partial_t$ is seen to be $\xi=(2 G M/r-1)/k^2$, which clearly can only vanish, and thus define a Killing
horizon, for $M>0$. Taking the asymptotic limit, one obtains $\lim_{r\to\infty}1/\sqrt{-\xi} = |k|$. Therefore, the absolute value of
the Dirac observable $k$ is related to the proper time of an asymptotic observer. However, the line element \eref{pre-Schwarzschild} is independent of the sign of $k$. This sign is
in fact related to a time inversion ($x\to-x$), since, from the evolution equation $r'=N k$, it is clear
that ${\rm sgn}(r')={\rm sgn}(k)$. Therefore, the sign of $k$ is simply encoding whether
the areal radius of the spheres of symmetry increases or decreases with $x$.

In order to see this more explicitly, in App.~\ref{app:null} we compute the expansions $\theta_{(i)}$ (with $i=1,2$) of the null vectors $n_i$ normal to the spheres of constant $x$ and $t$
for horizon-penetrating coordinates with $s\neq 0$. As commented above, we are particularly
interested in such coordinates in order to perform a quantization that simultaneously encodes
both sides of the horizon. In summary, the expansions of such vectors are defined as
\begin{equation}
 \theta_{(i)}:=\frac{1}{\sqrt{{\rm det}(q)}}n_i^a \nabla_a\left(\sqrt{{\rm det}(q)}\right),
\end{equation}
with $\sqrt{{\rm det}(q)}=r^2 \sin\theta$ being the volume element given by the determinant of the induced metric on the spheres. The expansions can be rewritten as
\begin{equation}
 \theta_{(i)}=\frac{2}{r}n^a_i \nabla_a r=\frac{2}{r}n^1_i r'(x)\,,
\end{equation}
and their signs are given by (cf. App.~\ref{app:null})
\begin{align}
 &{\rm sgn}(\theta_{(1)})=- {\rm sgn}(s\, k),\\
 &{\rm sgn}(\theta_{(2)})={\rm sgn}(s\, k(r-2GM)),
\end{align}
such that ${\rm sgn}(\theta_{(1)}\theta_{(2)})=-{\rm sgn}(r-2GM)$. Hence, the region $r>2GM$ is nontrapped, while $r<2GM$ is trapped (to the future) if $s\,k>0$ and antitrapped (trapped to the past) if $s\,k<0$.
Therefore, for horizon-penetrating charts with $s\neq 0$ and $M>0$,
the sign of the product $s\, k$ encodes whether the trapped
region is trapped to the future or to the past.
In this way, for a given sign of $s$, switching the sign of $k$ maps a trapped (black-hole) region to an antitrapped (white-hole) region or vice versa.
Note that, for $M\leq 0$, there is no trapped region as ${\rm sgn}(\theta_{(1)}\theta_{(2)})=-1$ everywhere.

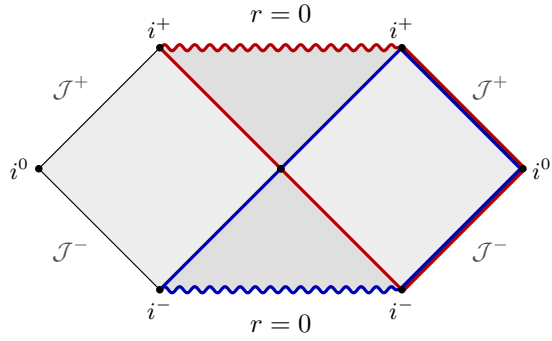
\begin{figure}[t]
\usetikzlibrary{decorations.pathmorphing, calc}
\center
\begin{tikzpicture}[scale=0.4]
\tikzset{
  horizonA/.style={red!70!black, very thick},
  horizonB/.style={blue!70!black, very thick},
  singularityA/.style={decorate, decoration={snake, amplitude=0.4mm, segment length=2mm}, very thick, red!70!black},
  singularityB/.style={decorate, decoration={snake, amplitude=0.4mm, segment length=2mm}, very thick, blue!70!black},
  scri/.style={font=\small, black!70},
  corner/.style={font=\small},
  vertex/.style={circle, fill=black, inner sep=1pt}
}

\node (I)    at ( 4,0)   {};
\node (II)   at (-4,0)   {};
\node (III)  at (0, 2.5) {};
\node (IV)   at (0,-2.5) {};

\path
  (II) +(90:4)  coordinate[label={[corner]90:$i^+$}]  (IItop)
       +(-90:4) coordinate[label={[corner]-90:$i^-$}] (IIbot)
       +(0:4)   coordinate                            (IIright)
       +(180:4) coordinate[label={[corner]180:$i^0$}] (IIleft)
       ;
\path
   (I) +(90:4)  coordinate[label={[corner]90:$i^+$}] (Itop)
       +(-90:4) coordinate[label={[corner]-90:$i^-$}] (Ibot)
       +(180:4) coordinate                            (Ileft)
       +(0:4)   coordinate[label={[corner]0:$i^0$}]   (Iright)
       ;

\fill[gray!14] (IIleft) -- (IItop) -- (IIright) -- (IIbot) -- cycle;
\fill[gray!14] (Ileft)  -- (Itop)  -- (Iright)  -- (Ibot)  -- cycle;

\fill[gray!25] (IItop) -- (IIright) -- (Itop) -- cycle;
\fill[gray!25] (IIbot) -- (IIright) -- (Ibot) -- cycle;

\draw[gray!60] (IIleft) -- (IItop) -- (IIright) -- (IIbot) -- (IIleft) -- cycle;
\draw[gray!60] (Ileft)  -- (Itop)  -- (Iright)  -- (Ibot)  -- (Ileft)  -- cycle;

\draw (IIleft) -- node[scri, midway, above left]  {$\mathcal{J}^+$} (IItop);
\draw (IIbot) -- node[scri, midway, below left]   {$\mathcal{J}^-$} (IIleft);

\draw (Itop) -- node[scri, midway, above right]   {$\mathcal{J}^+$} (Iright);
\draw (Iright) -- node[scri, midway, below right] {$\mathcal{J}^-$} (Ibot);

\draw[horizonA] (Ibot) -- (Ileft) -- (IItop);
\draw[horizonB] (IIbot) -- (Itop);

\draw[horizonA] ($(Itop)+(0.04,0.04)$)   -- ($(Iright)+(0.04,0.04)$);
\draw[horizonB] ($(Itop)+(-0.04,-0.04)$) -- ($(Iright)+(-0.04,-0.04)$);
\draw[horizonA] ($(Iright)+(0.04,-0.04)$) -- ($(Ibot)+(0.04,-0.04)$);
\draw[horizonB] ($(Iright)+(-0.04,0.04)$) -- ($(Ibot)+(-0.04,0.04)$);

\draw[singularityA] (IItop) -- (Itop)
      node[midway, above=2mm, black] {$r=0$};
\draw[singularityB] (IIbot) -- (Ibot)
      node[midway, below=2mm, black] {$r=0$};

\foreach \p in {IItop,IIbot,IIleft,IIright,Itop,Ibot,Ileft,Iright}
  \node[vertex] at (\p) {};

\end{tikzpicture}
\caption{The boundary of the black-hole domain (ingoing Eddington-Finkelstein chart) is highlighted
in red, while the boundary of the white-hole domain (outgoing Eddington-Finkelstein chart)
is drawn in blue.
The present minisuperspace formulation can describe the spacetime
region corresponding to either of these domains, but it can not capture
the complete maximal extension of the spacetime.
}
\label{penrosediagram}
\end{figure}

It turns out that, in this minisuperspace approach, it is not possible to construct a chart that covers the complete
maximal extension of the Schwarzschild black hole as given, for instance, by the Kruskal-Szekeres coordinate chart. Instead, we are restricted to charts with a domain given by half of the Kruskal-Szekeres diagram, such as ingoing or outgoing Eddington--Finkelstein charts
(see Fig.~\ref{penrosediagram}).

\subsection{Gravitational observables}

The Misner--Sharp mass \eref{mf-gen}--\eref{mf} introduced above is a scalar quantity that encodes the mass content of the spacetime. It is also a weak Dirac observable, and thus an invariant. However, not all spacetime scalars defined from the phase-space variables are (weak) Dirac observables. It is thus important to examine the relational observables associated to other scalars that describe the geometry and gravitational pull of the spacetime.

For a spherically symmetric metric there are only two scalars that encode information about spacetime curvature, namely, the four-dimensional Ricci scalar $R$ and the Newman--Penrose scalar $\Psi_2$. In terms of the canonical variables they read
\begin{align}\label{ricci}
R &=-\frac{1}{r^2}\left({2C}+\frac{3up_u'+rp_r'}{NM_0}\right),\\\label{psi2}
   \Psi_2 &=-\frac{1}{r^3}\left({G}M+{\frac{r}{3}C}+\frac{3urp_u'-r^2p_r'}{12NM_0}\right).
\end{align}
Clearly, when the equations of motion \eref{eom} are imposed, we have
\begin{align}\label{eq:ronshell}
R&=0\,,\\ \label{eq:psionshell} \Psi_2&=-\frac{{G}M}{r^3}\,.
\end{align}
Using the construction of relational observables discussed in Sec. \ref{sec:clrelob}, it is straightforward to verify that the corresponding observables are simply
\eq{
\Ob_R&= 0 \,,\\
\Ob_{\Psi_2}&=-\frac{GM}{\ell^3 e^{3\sigma}}\,.
}

The causal structure of the spacetime is encoded in the squared norm of the Killing field. For convenience, we repeat its expression here:
\begin{equation}
    \xi=-\frac{u}{r}.
\end{equation}
The corresponding relational observable is given by
\begin{equation}
    \mathcal{O}_\xi=-\frac{1}{k^2}\left(1-\frac{2GM}{\ell e^\sigma}\right).
\end{equation}

Other interesting scalars that encode information about the gravitational pull of the black hole,
are the proper acceleration of stationary observers and the surface gravity at the horizon. Consider a stationary observer on a static region $(\xi>0)$ that moves along the timelike Killing field $\partial_t$ with constant values of $(x,\theta,\phi)$. The corresponding four-velocity reads
\begin{equation}
 u^a=\frac{(\partial_t)^a}{\sqrt{-\xi}}=\sqrt{\frac{r}{u}}(\partial_t)^a \,,
\end{equation}
while the four-acceleration is given by
\begin{equation}
 u^b\nabla_b u^a=\frac{1}{2N^2}\left(\frac{u}{r}\right)'\left[\frac{sr}{u}(\partial_t)^a+(\partial_x)^a\right] \,.
\end{equation}
The norm of this vector defines the hovering acceleration $a$ needed for this observer to remain stationary,
\begin{equation}\label{classica}
 a^2:=u^b u^c(\nabla_b u_a)(\nabla_c u^a)=\frac{r}{4 N^2 u}\left[\left(\frac{u}{r}\right)'\right]^2=-\frac{(\xi')^2}{4 N^2\xi},
\end{equation}
which diverges as the observer approaches the Killing horizon, where $\xi=0$. However, one can define a ``regularized acceleration'' $\alpha$ as
\begin{equation}\label{classicalpha}
 \alpha^2:= -\xi a^2=\left(\frac{\xi'}{2 N}\right)^2,
\end{equation}
which remains finite at the Killing horizon. In fact, up to a normalization constant, $\alpha$ reproduces the surface gravity $\kappa$ at the horizon, which can be defined as
\begin{equation}\label{classickappa}
 \kappa^2= \lim_{r\rightarrow r_h} \frac{\alpha^2}{-\xi_0}\\
=\lim_{r\rightarrow r_h}\frac{1}{-\xi_0}\left(\frac{\xi'}{2N}\right)^2\,,
\end{equation}
with $\xi_0$ being the value of $\xi$ in the asymptotic limit $\xi_0:=\lim_{r\to\infty}\xi$. Note that, if the Killing vector is normalized at infinity, i.e. $\xi_0\equiv-1$, the regularized acceleration evaluated at the horizon coincides with the surface gravity, $\kappa^2=\lim_{r\to r_h}\alpha^2$. In terms of the phase-space variables, we can write
\begin{equation}\label{eq:regacc}
\alpha^2=\frac{1}{4 r^4 M_0^2}\left(up_u-rp_r\right)^2\,,
\end{equation}
and the relational observable corresponding to $\alpha^2$ reads
\begin{equation}\label{classicalkillingnorm}
\Ob_{\alpha^2}=
\frac{G^2M^2}{k^2\ell^4e^{4\sigma}}\,.
\end{equation}

The spacetime scalars $R$, $\Psi_2$, and $\alpha^2$
are the objects that will be promoted to quantum operators in the next setion,
in order to interpret the quantum properties of the geometric structures.
Even if these objects will be well defined operators on the Hilbert space,
the geometrical interpretation of their expectation values should
only be considered appropriate when describing a scenario with
a slightly deformed classical geometry. We will formalize this idea in more
detail below.

\section{Quantum theory}\label{sec:QuanTh}

In the quantum theory, only the quantum state and quantum operators are directly meaningful,
while, as commented above, the notion of a spacetime geometry is something that may be inferred only in the classical limit.
Furthermore, in order to avoid gauge ambiguities in the interpretation of quantum
effects, we must consider quantum Dirac observables. In what follows, we describe their construction
as symmetric operators in the physical Hilbert space.

\subsection{Kinematical Hilbert space and basic operators}\label{sec:KinH}

Our starting point for the quantization of the model is to select a set of elementary functions on the classical phase space and represent them as self-adjoint operators acting on a suitable kinematical Hilbert space, in such a way that Poisson brackets are mapped to commutators \cite{Dirac1}. The canonical pair $(u,p_u)$ can be represented on $L^2(\mathbb{R},du)$, with $\hat{u}$ acting by multiplication and $\widehat{p}_u:=-i\hbar\,\partial_u$ by differentiation.
These operators are essentially self-adjoint and satisfy the canonical commutation relation $\left[\hat{u},\widehat{p}_u\right]=i\hbar$ in a common dense domain.
However, the same construction cannot be applied to the pair $(r,p_r)$. Since $r$ is classically restricted to the positive half-line,
the naive quantization of $p_r$ as the momentum operator
$
\widehat{p}_r= -i\hbar\,\partial_r
$
in the Hilbert space $L^2(\mathbb{R}^+,{\rm d}r)$
fails to be self-adjoint. We instead represent the classical quantity $r p_r$ by the operator
\begin{equation}
    \widehat{rp}_r:=-i\hbar (r\partial_r+\gamma+1)\,.
\end{equation}
It can be shown that $\widehat{rp}_r$ generates a unitary representation of the dilation group on $L^2(\mathbb{R}^+,r^{2\gamma+1}dr)$, and therefore it is self-adjoint \cite{vilenkin} (see App.~\ref{app:dilationgroup}). We will see below that, although $\gamma$ represents a freedom in the representation of this group, the physical results will not depend on it. If we further introduce $\hat{r}$ as a multiplication operator, it follows that $[\hat{r},\widehat{rp}_r]=i\hbar \, \hat{r}$ in a common dense domain, providing the quantum analog of the classical relation $\{r,rp_r\}=r$. The  kinematical Hilbert space of the system is thus given by the tensor product
\begin{equation}
    \mathcal{H}_{kin}:=L^2(\mathbb{R},du)\otimes L^2(\mathbb{R}^+,r^{2\gamma+1}dr)\,.
\end{equation}
The normalization and the completeness relation of the (generalized) basis vectors $\ket{u,r}$ read
\eq{\label{eq:ortho-complete}
    \langle u,r|u',r'\rangle&=r^{-2\gamma-1}\delta(u-u')\delta(r-r')\,,\\
    \widehat{I} &=\int_0^{\infty}\D r\,r^{2\gamma+1}\int_{-\infty}^{\infty}\D u\,\ket{u,r}\!\!\bra{u,r}\,,
}
where $\langle{\cdot|\cdot}\rangle$ is the kinematical inner product and $\widehat{I}$ is the identity operator on the kinematical Hilbert space.

As explained above, the constraint $C={p_u p_r}/{M_0^2}-1$
defines the Hamiltonian as $H=NM_0\, C$ and completely
encodes the classical dynamics of the model.
Following Dirac's quantization procedure \cite{Dirac1},
one needs to promote $C$ to an operator and define
the physical quantum states as those annihilated by
the constraint $\widehat C \ket{\Psi}=0$. However, 
in the expression for $C$ the momentum $p_r$
appears and, as explained above, we can not define
a corresponding self-adjoint operator. Therefore,
instead of $C$, we consider the rescaled constraint
${\cal C}:=r C$ to be promoted to an operator.
We note that classically $C$ and ${\cal C}$ define
exactly the same dynamics as both constraint equations $C=0$ and ${\cal C}=0$ are equivalent,
and for any basic phase-space function $f$ one
has $f'=NM_0 \left\{f,C \right\}=NM_0/r \left\{f,{\cal C} \right\}$
on shell. In this way, we introduce the self-adjoint constraint operator
\begin{equation}\label{eq:constraintop}
    \widehat{\cal C}:=-\left(\hbar/M_0\right)^2\partial_u(r\partial_r+\gamma+1)-r \,,
\end{equation}
where we have considered a symmetric ordering between
$\widehat r$ and $\widehat{r p_r}$,
and the quantum analogue of Eq.~\eref{constraint}
is thus given by the eigenvalue equation $\widehat{\cal C}\ket{\Psi} = 0$.

We can now construct the quantum
operators corresponding to different classical quantities of
interest from their expression in terms of the basic variables
$u$, $r$, $p_u$, $r p_r$, and ${\cal C}$. The only freedom left corresponds
to the operator orderings, and, as done above, we will choose a symmetric ordering in the basic operators. In particular, from \eqref{eq:mf}, the Misner-Sharp mass operator reads,
\eq{
\widehat{M}&=\frac{1}{2G}\left[\frac{\hbar^2}{M_0^2}\left(u\partial_u^2+\partial_u\right)+r\right]\,.
}
In order to construct the operators corresponding to the
curvature scalars defined in \eqref{ricci}--\eqref{psi2},
we proceed as follows. On the one hand, we replace $C$ by ${\cal C}/r$,
which can be promoted to an operator straightforwardly.
On the other hand, for the time derivatives of the momenta,
we consider that classically $p_u'/N=\{p_u,C\}=\{p_u,{\cal C}/r\}$,
and
$p_r'/N=((r p_r)'- p_r r')/(N r)=\{r p_r,{\cal C}/r\}/r-r p_r p_u/(M_0 r^2)$. On this basis, we construct the corresponding quantum operators by replacing Poisson brackets with commutators and applying a symmetric ordering of the resulting expressions.
All in all, this quantization procedure leads
to the following curvature operators:
\eq{\label{eq:kincurv}
\widehat{R}&=\frac{1}{2r^3}\left[\frac{\hbar^2}{M_0^2}\left(2r\partial_u\partial_r+(2\gamma-1)\partial_u\right)+2r\right]\,,\\
\widehat{\Psi}_2&=\frac{1}{6r^3}\left[\frac{\hbar^2}{M_0^2}\left(3u\partial_u^2-2r\partial_u\partial_r-2(\gamma-2)\partial_r\right)+r\right]\,.
}
Finally, choosing a symmetric ordering for the basic operators,
from \eqref{eq:regacc} the squared of the regularized acceleration
of a hovering observer is represented by the operator,
\begin{equation}\label{eq:kinregacc}
\hat{\alpha}^2= -\frac{\hbar^2}{4r^4M_0^2}\bigg[u^2\partial_u^2-2ur\partial_u\partial_r+r^2\partial_r^2-2(\gamma-2)u\partial_u+2(\gamma-1)r\partial_r+\frac{1}{6}
\Big(17 + 6 (\gamma-3) \gamma\Big)\bigg]\,,
\end{equation}
while the squared norm of the Killing field \eqref{eq.killing} is promoted to
\begin{equation}\label{eq:kinkilling}
\widehat{\xi}=-\hat{u}\circ\widehat{r^{-1}},
\end{equation} 
where $\circ$ denotes the composition of operators.

\subsection{Physical Hilbert space}
\label{sec:physHS}

The physical Hilbert space can be constructed as a vector space of solutions to the quantum constraint equation $\widehat{\cal C}\ket{\Psi} = 0$ endowed with an induced inner product.
Since $\hat{p}_u$ commutes with $\widehat{\cal C}$ off shell, it is possible to construct their simultaneous eigenstates,
\eq{\label{eq:constraint-eigenstates}
    \psi_{\lambda,k}(u,r):=\left\langle u,r|\lambda,k\right\rangle =\frac{M_0}{2\pi\hbar}\,|k|^{-\frac{1}{2}}\,r^{-\gamma-1+\frac{\I\lambda M_0}{\hbar k}}e^{\frac{\I M_0}{\hbar}\left(ku+\frac{1}{k}r\right)}\,,
}
such that,
\begin{align}
\widehat{\cal C}\, \ket{\lambda,k}&=\lambda\, \ket{\lambda,k},\\
\widehat{p}_u\, \ket{\lambda,k}&=M_0\, k\, \ket{\lambda,k},
\end{align}
with $\lambda\in\mathbb{R}$ and $k\in\mathbb{R}\setminus\{0\}$.
The eigenstates $\ket{\lambda,k}$
provide a complete basis
of vectors for the kinematical Hilbert space.
By construction, they are orthonormal with respect to the kinematical inner product,
\begin{equation}\label{kininnerproduct}
    \left\langle \lambda',k'|\lambda,k\right\rangle=\delta(\lambda'-\lambda)\delta(k'-k) \,,
\end{equation}
and fulfill the completeness relation
\eq{
\int_{-\infty}^{\infty}\D k\int_{-\infty}^{\infty}\D\lambda\,\ket{\lambda,k}\!\!\bra{\lambda,k} = \widehat{I} \,.
}
We note that any superposition of the vectors $|k\rangle:=|\lambda=0,k\rangle$ is a solution to the quantum constraint equation, $\widehat{\cal C}\ket{\Psi} = 0$. Therefore, in order to construct the physical Hilbert space,
from \eqref{kininnerproduct}, we define the physical (or induced)
inner product $(\lambda,k|\lambda, k')$ as
\begin{equation}
\delta(\lambda'-\lambda)(\lambda,k'|\lambda,k):=\left\langle \lambda',k'|\lambda,k\right\rangle\,,
\end{equation}
that is, $(\lambda,k'|\lambda,k):=\delta(k'-k)$,  which is a well-defined expression even for $\lambda'=\lambda=0$. In this way, physical states are given by the superpositions $\ket{\Psi} = \int_{\mathbb{R}}\mathrm{d}k\,\Psi(k)\ket{k}$ that are normalizable with respect to the physical inner product $(\cdot|\cdot)$.\footnote{The $\ket{k}$ states are, strictly speaking, only defined for $k\neq0$, but one can consider the superposition $\ket{\Psi} = \int_{\mathbb{R}}\mathrm{d}k\,\Psi(k)\ket{k}$ for $k\in\mathbb{R}$ by defining the states $\ket{k = 0}\equiv0$. From now on, we denote the integration range of the $k$ label as the real line.}
The vector space of these states, equipped with the physical inner product,
is the physical Hilbert space. More precisely, given two such physical states
$\ket{\Psi_1} = \int_{\mathbb{R}}\mathrm{d}k\,\Psi_1(k)\ket{k}$
and $\ket{\Psi_2} = \int_{\mathbb{R}}\mathrm{d}k\,\Psi_2(k)\ket{k}$,
their physical inner product reads explicitly
\begin{equation}
(\Psi_1|\Psi_2)=\int_\mathbb{R} {\rm d}k\,\Psi_1^*(k)\Psi_2(k).
\end{equation}
Thus, the normalizability condition implies suitable boundary (fall-off) conditions for the wave function $\Psi(k)$ as $k\to\pm\infty$.

Finally, we note that $k$ is simply a label on physical wave functions,
which corresponds to the eigenvalue of the Dirac observable $\widehat{p}_u$.
As discussed in Sec.~\ref{sec.trapped}, in the classical model, $1/k^2$
corresponds to the asymptotic limit of the squared norm of the
Killing field. Thus the absolute value $|k|$ is related to the proper time of
the asymptotic observer, while its sign is related to the direction
of time and trapping features.
More precisely, the sign of the combination $(k\, s)$ encodes the trapping
properties of the classical chart under consideration.
However, $s$ does not appear in the quantum model, and thus the only
trace of that combination is the label $k$ of the physical states.
Therefore, although the sign of $k$ has no intrinsic meaning,
one may interpret a sign switch in $k$ (assuming $s$ is kept fixed)
as changing from a trapped (black-hole) to an antitrapped (white-hole)
region. In any case, since both types of chart admit
the same quantum description, in what follows
we will generally use the terminology ``black hole'' for convenience,
without implying a distinction at the quantum level between
the trapped and antitrapped cases.

\subsection{Gauge fixing and quantum relational observables}

In the previous subsection, the physical inner product $(\cdot,\cdot)$ has been defined from the kinematical product $\left\langle \lambda',k'|\lambda,k\right\rangle=:\delta(\lambda'-\lambda)(\lambda,k'|\lambda,k)$ in an implicit manner, by extracting the delta function $\delta(\lambda'-\lambda)$ that becomes divergent for physical states
with $\lambda = 0$ \cite{Ashtekar:94,Landsman:95,Marolf:97,Embacher:98,Giulini:99,Giulini:99-2,Giulini:2000,Marolf:2000,Halliwell,Chataignier:2019kof,Chataignier:2020fys,Chataig:Thesis}. Alternatively, it can also be defined explicitly by the insertion of a `gauge-fixing operator' in the kinematical inner product. To see this, we follow Refs.~\cite{Chataignier:2019kof,Chataignier:2020fys,Chataig:Thesis} and we promote the classical
gauge-fixing function $\chi$ to an operator $\widehat{\chi}$ that is self-adjoint with respect to the kinematical inner product.

As in the classical theory, we consider the gauge-fixing function $\widehat{\chi} = \ln(\hat{r}/\ell)$ and introduce the projector
\eq{
\widehat{P}_{\chi = \sigma}:= \int_{-\infty}^{\infty}\D \chi\,\int_{-\infty}^{\infty}\D u\,(\ell e^\chi)^{2(\gamma+1)}\,\delta(\chi-\sigma)\ket{u,\chi}\!\!\bra{u,\chi} = \int_{-\infty}^{\infty}\D u\,(\ell e^\sigma)^{2(\gamma+1)}\,\ket{u,\chi = \sigma}\!\!\bra{u,\chi = \sigma}\,,\label{eq:projector}
}
where
$\ket{u,\chi}:=\ket{u,r=\ell e^\chi}$ are eigenstates of $\widehat{\chi}$ obeying the orthonormality and completeness relations,
\eq{\label{eq:ortho-complete-rho}
    \langle u,\chi|u',\chi'\rangle&=(\ell e^\chi)^{-2(\gamma+1)}\delta(u-u')\delta(\chi-\chi')\,,\\
    \widehat{I} &=\int_{-\infty}^{\infty}\D \chi\,\int_{-\infty}^{\infty}\D u\,(\ell e^\chi)^{-2(\gamma+1)}\,\ket{u,\chi}\!\!\bra{u,\chi}\,.
}
Using Eqs. \eref{constraint-eigenstates} and \eref{projector}, we find the matrix elements
\eq{
\braket{\lambda,k'|\widehat{P}_{\chi = \sigma}|\lambda,k} = \frac{M_0}{2\pi\hbar}|k|^{-1}\delta(k'-k) \ . 
}
As can be seen, these matrix elements are independent of the
parameter $\gamma$. In addition, we
see that $\widehat{P}_{\chi = \sigma}$ can be inverted on the physical Hilbert space. More precisely, its inverse reads
\eq{
\widehat{\Delta}_{\chi} := \frac{2\pi\hbar}{M_0}\int_{-\infty}^{\infty}\D k\int_{-\infty}^{\infty}\D\lambda\, |k|\ket{\lambda,k}\!\!\bra{\lambda,k} \,.
}
Therefore, we obtain the identity
\eq{\label{eq:gf-ip}
(k'|k) := \delta(k'-k) = \braket{k'|\widehat{\Delta}^{\frac12}_{\chi}\widehat{P}_{\chi = \sigma}\widehat{\Delta}^{\frac12}_{\chi}|k} \,,
}
with
\eq{\label{eq:QuantumFad}
\widehat{\Delta}^{\frac12}_{\chi} := \sqrt{\frac{2\pi\hbar}{M_0}}\int_{-\infty}^{\infty}\D k\int_{-\infty}^{\infty}\D\lambda\, |k|^{\frac12}\ket{\lambda,k}\!\!\bra{\lambda,k}
}
being an invariant operator (that is, it commutes with $\widehat{\cal C}$) that plays the role of a quantum version of the square-root of the classical Faddeev--Popov determinant $\Delta_\chi(\sigma)$.

We can thus interpret Eq.~\eref{gf-ip} as defining the physical inner product in terms of a gauge-fixing operator $\widehat{\Delta}^{\frac12}_{\chi}\widehat{P}_{\chi = \sigma}\widehat{\Delta}^{\frac12}_{\chi}$ inserted on the kinematical inner product. Likewise, it can be seen as defining the matrix elements of the relational observable that corresponds to the identity. More precisely, given a kinematical self-adjoint operator $\hat{f}$, we define the associated quantum relational observable relative to the gauge condition $\chi = \sigma$ as follows,
\eq{
\widehat{\Ob}_f\equiv \widehat{\mathcal{O}}[f|\chi=\sigma]&:= \frac{1}{2}\int_{-\infty}^{\infty}\D k{\int_{-\infty}^{\infty}\D k'}\int_{-\infty}^{\infty}\D\lambda\, \braket{\lambda,k'|\widehat{\Delta}^{\frac12}_{\chi}(\hat{f}\widehat{P}_{\chi = \sigma}+\widehat{P}_{\chi = \sigma}\hat{f})\widehat{\Delta}^{\frac12}_{\chi}|\lambda,k}\ket{\lambda,k'}\!\!\bra{\lambda,k} \,.
}
Using Eq. \eref{QuantumFad} this expression can be written as
\eq{\label{observablef}\widehat{\Ob}_f= \frac{\pi\hbar}{2M_0}\int_{-\infty}^{\infty}\D k{\int_{-\infty}^{\infty}\D k'}\int_{-\infty}^{\infty}\D\lambda\, |k'k|^{\frac12}\braket{\lambda,k'|\hat{f}\widehat{P}_{\chi = \sigma}+\widehat{P}_{\chi = \sigma}\hat{f}|\lambda,k}\ket{\lambda,k'}\!\!\bra{\lambda,k} \,.}
In particular, if $\hat{f} = \widehat{I}$, we trivially obtain $\widehat{\Ob}_I = \widehat{I}$, as expected.  
Given a physical state $\ket{\Psi} = \int_{\mathbb{R}}\mathrm{d}k\,\Psi(k)\ket{k}$, the expectation value of a relational observable calculated with respect to the physical inner product is thus given by
\eq{\label{eq:general-expval}
\braket{\widehat{\Ob}_f}:=(\Psi|\widehat{\Ob}_f\Psi) =\frac{2\pi\hbar}{M_0}\,\mathfrak{Re}\int_{-\infty}^{\infty}\D k'\int_{-\infty}^{\infty}\D k\, \Psi^*(k')\Psi(k)|k'k|^{\frac12}\braket{k'|\hat{f}\widehat{P}_{\chi = \sigma}|k}\,,
}
where $\mathfrak{Re}$ denotes the real part of a complex number. Note that, since the matrix elements of $\widehat{\Delta}_\chi^{\frac{1}{2}}$ and $\widehat{P}_{\chi=\sigma}$ do not depend on $\gamma$, $\widehat{\mathcal{O}}_f$ will be independent of $\gamma$ whenever the matrix elements of $\hat f$ are as well. It can be shown that this is indeed the case for the kinematical operators constructed in Section~\ref{sec:KinH}.

We can now use the general expression \eqref{observablef} to obtain the observables corresponding
to the basic kinematical operators
$\hat{u}$, $\hat{p}_{u}$, $\hat{r}$, and $\widehat{rp}_r$.
Their form is easier to be expressed as their action on the physical wave function $\Psi(k)$:
\eq{
\widehat{\mathcal{O}}_u\Psi(k) &=  \frac{i\hbar}{M_0}\Psi'(k)+\frac{\ell e^\sigma}{k^2}\Psi(k)\,,\\
\widehat{\mathcal{O}}_r\Psi(k) &= \ell e^\sigma\Psi(k)\,,\\
\widehat{\mathcal{O}}_{p_u}\Psi(k)&=M_0k\,\Psi(k)\,,\\
\widehat{\mathcal{O}}_{rp_r}\Psi(k)&= \frac{\ell M_0e^\sigma}{k}\Psi(k)\,.
}

Their expectation values on physical states $|\Psi\rangle=\int {\rm d}k\,\Psi(k) |k\rangle$ can then be computed either by using \eqref{eq:general-expval}
or directly from these last expressions as $\langle\widehat{\cal O}_f\rangle:=(\Psi|\widehat{\cal O}_f\Psi)=\int {\rm d}k\Psi^*(k) \widehat{\cal O}_f \Psi(k)$.
In particular, since the gauge-fixing function has
been chosen as a function of $r$, for any physical state
the expectation value of the quantum relational observable $\widehat{\mathcal{O}}_r$
simply reproduces the classical relation (\ref{rel.radius}):
\begin{equation}
\langle \widehat{{\cal O}}_r \rangle=\ell e^\sigma,
\end{equation}
with the parameter $\sigma$ in the range $\sigma\in(-\infty,+\infty)$.
The classical asymptotically flat infinity
corresponds to $\sigma\to+\infty$, while the classical singularity is located at $\sigma\to-\infty$.

Note also that the quantum relational observable $\widehat{\mathcal{O}}_{p_u}$  associated to the Dirac observable $\widehat{p}_u$
is independent of the variable $\sigma$, and thus it is constant through evolution.
The same applies to other Dirac observables, such as the mass observable,
\begin{equation}\label{eq.Om}
\widehat{\mathcal{O}}_M\Psi(k)=-\frac{i\hbar}{2GM_0}
\left[
k^{2}\,\Psi'(k)
+
k\,\Psi(k)
\right]
\,.
\end{equation}

In the following, we will construct the quantum operators
associated to classical spacetime scalar quantities.
First, the kinematical scalar curvature operators \eqref{eq:kincurv}
lead to the observables:
\begin{align}
    \label{eq:relRopp}\widehat{\mathcal{O}}_R\Psi(k) &=0,\\
    \label{eq:relPSopp}\widehat{\mathcal{O}}_{\Psi_2}\Psi(k)&=-\frac{G}{\ell^3 e^{3\sigma}}\widehat{\mathcal{O}}_M\Psi(k).
\end{align}
From here we can see that their expression is formally equivalent
to their classical on-shell values, as given by Eqs.~\eqref{eq:ronshell} and \eqref{eq:psionshell}, respectively.
Therefore, for any physical
quantum state, the observable corresponding to the Ricci scalar will
be vanishing, and thus also its expectation value and all higher-order
moments, i.e., $\langle \widehat{{\cal O}}_R^n\rangle=0$ for any integer $n\geq 1$.
Concerning the observable associated to $\Psi_2$, its expectation
value is simply $\langle \widehat{{\cal O}}_{\Psi_2}\rangle=- G \langle{\widehat{\cal O}}_M\rangle/\langle{\widehat{\cal O}_r}\rangle^3$.
We note that, at the classical singularity $\langle{\widehat{\cal O}}_r\rangle\to 0$, generically
$\langle \widehat{{\cal O}}_{\Psi_2}\rangle$ diverges, unless
$\langle{\widehat{\cal O}}_M\rangle=0$ identically.

Concerning observables that encode the causal structure of the spacetime and its gravitational pull, on the one
hand,
the quantum observable representing the squared norm
of the Killing \eqref{eq:kinkilling} is given by 
\begin{equation}
    \label{eq:relkillopp}\widehat{\mathcal{O}}_\xi=-\frac{\widehat{\mathcal{O}}_u}{\ell e^\sigma},
\end{equation}
and thus its expectation value is just $\langle \widehat{\mathcal{O}}_\xi \rangle=-\langle \widehat{\mathcal{O}}_u \rangle/\langle \widehat{\mathcal{O}}_r \rangle$.
At the level of expectation values, that is, disregarding fluctuations,
we will interpret the sign of $\langle \widehat{\mathcal{O}}_\xi \rangle$
to encode the causal character of the Killing field, and thus whether
for a given value of $\sigma$ the state describes a homogeneous
or a static spacetime region.
In this way, its vanishing will define the position
of the quantum Killing horizon, i.e., the location of the horizon
will be given by $\langle\widehat{\mathcal{O}}_r\rangle|_{\sigma=\sigma_h}$ where $\sigma_h$ is defined by the equation
$\langle \widehat{\mathcal{O}}_\xi \rangle\big|_{\sigma=\sigma_h}=0$.
However, this interpretation will only be valid for semiclassical
states with negligible fluctuations.

On the other hand, the quantum observable representing the squared norm
of the regularized acceleration  \eqref{eq:kinregacc} reads 
\begin{equation}
    \widehat{\mathcal{O}}_{\alpha^2}\Psi(k)=\frac{\hbar^2}{4M_0^2\ell^4e^{4\sigma}}\left[-k^2\Psi''(k)-2k\Psi'(k)+2\Psi(k)\right].
\end{equation}

Making use of the expectation values $\langle\widehat{\mathcal{O}}_{\alpha^2}\rangle$
and $\langle\widehat{\mathcal{O}}_\xi\rangle$,
we introduce the effective hovering  acceleration for a static observer,
\begin{equation}\label{eq.acceleration}
 a^2_\text{eff}:=\frac{\langle\widehat{\mathcal{O}}_{\alpha^2}\rangle}{\langle\widehat{\mathcal{O}}_\xi\rangle},
\end{equation}
and the effective surface gravity of the quantum horizon,
\begin{equation}\label{eq.kappa}
\kappa_\text{eff}^{2}:=\frac{\lim _{\sigma\rightarrow \sigma_h}\langle\widehat{\mathcal{O}}_{\alpha^2}\rangle}{\lim_{\sigma\rightarrow \infty}\langle\widehat{\mathcal{O}}_\xi\rangle}.
\end{equation}
These quantities are effective in the sense that they are not defined as the expectation value of an operator.
Instead, they are defined from their classical expressions \eqref{classicalpha}--\eqref{classickappa} in terms of $\xi$ and $\alpha^2$. This is because constructing the quantum relational operators associated to the acceleration and the surface gravity from some kinematical operator is cumbersome. In the case of the acceleration, one has to deal with inverse powers of $\hat{u}$,  
which are problematic when inserted into the equation defining the quantum relational observables \eqref{observablef}.
Concerning the surface gravity, the problem is that the corresponding classical expression
is itself nonlocal, since it considers the value of $\alpha^2$ at the horizon and the value of $\xi$ at spatial infinity,
which makes difficult even to define the corresponding kinematical operator.
In a similar way as the notion of the horizon, the interpretation of these effective quantities
as describing the pull of the black hole will only be reasonable for
highly semiclassical states.

\section{Construction of relevant quantum states}\label{sec:states}

In this section, we construct specific physical states that could be
suitable to provide a semiclassical picture of a classical geometry.
More precisely, to gain a deeper understanding of the nature of the physical quantum states,
in Sec.~\ref{sec.massstates} we study the eigenstates of the
(relational) mass operator ${\widehat{\cal O}}_M$. These eigenstates will be useful
to obtain information about the probability distribution of the mass
observable in more general states. Then, in Sec.~\ref{sec.semclstates},
we construct the so-called intelligent states by requiring them to saturate the Robertson–Schrödinger uncertainty
relation for the quantum relational observables $\widehat{\mathcal O}_{p_u}$ and $\widehat{\mathcal O}_M$.
This condition will define a four-parameter family of states. Among these, we will identify the semiclassical
intelligent states as those with small relative fluctuations, and, in Sec.~\ref{sec.physicsintelligentstates},
we will study in detail their physical implications.

\subsection{Mass eigenstates}\label{sec.massstates}

As explained in Section \ref{sec:physHS}, physical quantum states can be expressed as superpositions of the form $\ket{\Psi} = \int_{\mathbb{R}}\mathrm{d}k\,\Psi(k)\ket{k}$ that are normalizable with respect to the physical inner product.
In this representation, $|\Psi(k)|^2$ provides the probability distribution
in $k$, which is the eigenvalue of the observable $\widehat{\Ob}_{p_u}$.
As explained above, classically $k$ is related to certain coordinate
choices, while the actual physical information of the solution is encoded
in the mass parameter $M$. Therefore, it is of high interest to
obtain the eigenstates of its corresponding observable $\widehat{\cal O}_M$,
such that we can write any given physical state as a superposition of mass
eigenstates and obtain its mass spectrum.

In order to construct the mass representation, let us thus determine the eigenstates $|M\rangle$ of the operator $\widehat{\Ob}_M$. For this purpose, we consider the  expansion 
\eq{
    |M\rangle=\int_{-\infty}^{\infty}\D k \,\Phi_M(k)|k\rangle.
}
Substituting this expansion into the eigenvalue equation,
\begin{equation}
\widehat{\cal O}_M | M\rangle= M | M\rangle,
\end{equation}
with $\widehat{\cal O}_M$ as defined in \eqref{eq.Om},
leads to a first-order differential equation for the wave function $\Phi_m(k)$:
\eq{-\frac{i\hbar}{2GM_0}\left(k^2\Phi_M'(k)+k\Phi_M(k)\right)=M\Phi_M(k).}
The solution for $k\neq 0$ is given by
\begin{equation}\label{eq.masseigenstates}
    \Phi_M(k)=\sqrt{\frac{GM_0}{\pi\hbar}}{|k|}^{-1}\,e^{-\frac{2iGM_0}{\hbar }\frac{M}{k}},
\end{equation}
where the normalization constant has been chosen such that these states form an orthonormal set with respect to the physical inner product, namely,
\eq{(M'|M)=\delta(M'-M).}

We note that, in principle, the mass eigenvalue $M$ is allowed to take arbitrary real values, including both positive and negative ones, as well as $M=0$.
The functional form of the eigenstates $\Phi_M(k)$ is essentially
given by a plane wave in $1/k$ with an amplitude moduled by $|k|^{-1}$.
Hence, the probability distribution $|\Phi_M(k)|^2$ does not
depend on the sign of $k$. However, the imaginary phase does depend on
the sign of $k$, and there is a symmetry $\Phi_{M}(k)=\Phi_{-M}(-k)$,
which may play an important role in the superposition of states
with positive and negative masses.

Strictly speaking,
any given mass eigenstate is not normalizable, and thus can not represent
a physical state. However, as they form a basis for the physical
Hilbert space, any physical state can be expressed as
\eq{\label{eq:massrep}|\Psi\rangle=\int \D M\,\widetilde{\Psi}(M) |M\rangle,}
where $\widetilde{\Psi}(M)=( M|\Psi)$ is the wave function in the mass representation.
Specifically, $|\widetilde{\Psi}(M)|^2$ provides information about the probability distribution of the relational mass observable
$\widehat{\Ob}_M$ for the quantum state under consideration.

\subsection{Intelligent states}
\label{sec.semclstates}

As already commented,
the classical Schwarzschild geometry is completely characterized
by the mass parameter $M$. In the previous subsection
we have constructed the mass eigenstates, but, as explained above, they are not
normalizable and, strictly speaking, a given mass eigenstate $|M\rangle$ can not represent a physical
state. In principle, a semiclassical state can be constructed by considering
a physical state of the form \eqref{eq:massrep} with $\widetilde{\Psi}(M)$ chosen
with a peaked profile around certain value $\bar M$. 
However, the issue is that there is a large amount of freedom in the construction
of this kind of semiclassical states, and one does not have a direct control
on the corresponding fluctuations. In addition, the canonical solution
of the equations of motion is parameterized not only by $M$, but also
by $k$, which in the quantum model corresponds to the eigenvalues of $\widehat{\mathcal{O}}_{p_u}$.

Therefore, in the following we will present
a natural family of physical semiclassical states for this model,
by considering
sharply peaked states around prescribed values of the corresponding
quantum relational observables $\widehat{\mathcal{O}}_M$ and $\widehat{\mathcal{O}}_{p_u}$,
and requesting them to be of minimal uncertainty in a specific sense.
For such a purpose, we will consider the Robertson-Schr\"odinger uncertainty
relation for these observables, which can be obtained by applying the Cauchy-Schwarz inequality
to the vectors $(\widehat{\mathcal{O}}_{p_u}
-\langle \widehat{\mathcal{O}}_{p_u}\rangle)|\Psi\rangle$ and $(\widehat{\mathcal{O}}_{M}-\langle \widehat{\mathcal{O}}_{M}\rangle)|\Psi\rangle$, and it reads
\begin{equation}
(\Delta{\mathcal{O}}_{p_u})^2\,(\Delta{\mathcal{O}}_M)^2\geq 
\frac{1}{4}|\langle[\widehat{\mathcal{O}}_{p_u},\widehat{\mathcal{O}}_{M}]\rangle|^2
+
\frac{1}{2}\left|\langle[\widehat{\mathcal{O}}_{p_u},\widehat{\mathcal{O}}_{M}]_+\rangle
-2\langle\widehat{\mathcal{O}}_{{p_u}}\rangle\langle\widehat{\mathcal{O}}_{M}\rangle\right|^2,\label{GHI}
\end{equation}
where $\Delta{\cal O}_f$ stands for the fluctuation of the corresponding
observable, $\Delta \mathcal{O}_f:=\sqrt{\langle\widehat{\Ob}_f^2\rangle-\langle\widehat{\Ob}_f\rangle^2}$,
and $[\cdot,\cdot]_+$ is the anticommutator.

Specifically, we want to construct a family of physical states,
\begin{equation}
    |\Psi\rangle=\int \D k\,\Psi(k)|k\rangle,
\end{equation}
such that $i/$ the expectation values of the Dirac observables
read $\langle \widehat{\mathcal{O}}_{p_u}\rangle=M_0\bar{k}$ and $\langle\widehat{\mathcal{O}}_M\rangle=\m0$
for certain chosen values $\bar k$ and $\bar M$,
and $ii/$ their fluctuations saturate the generalized Robertson-Schr\"odinger inequality \eqref{GHI}, that is,
\begin{equation}
(\Delta{\mathcal{O}}_{p_u})^2\,(\Delta{\mathcal{O}}_M)^2=
\frac{1}{4}|\langle[\widehat{\mathcal{O}}_{p_u},\widehat{\mathcal{O}}_{M}]\rangle|^2
+
\frac{1}{2}\left|\langle[\widehat{\mathcal{O}}_{p_u},\widehat{\mathcal{O}}_{M}]_+\rangle
-2\langle\widehat{\mathcal{O}}_{{p_u}}\rangle\langle\widehat{\mathcal{O}}_{M}\rangle\right|^2.
\end{equation}
These states are typically referred to as intelligent states.

The Cauchy-Schwarz inequality is saturated if and only if the corresponding vectors are collinear \cite{Kreyszig,ReedSimon}. Therefore,
considering the expectation values of the observables as indicated in the condition {\it i/},
the saturation property
{\it ii/} for the state $\ket{\Psi}$
implies that
\begin{equation}
    (\widehat{\mathcal{O}}_{p_u}
-M_0 \bar{k})|\Psi\rangle=-i \frac{2 G M_0^2}{\hbar w}(\widehat{\mathcal{O}}_M-\m0)|\Psi
\rangle,\end{equation}
for some complex constant $w=w_r+ i w_i$ with $w_r, w_i\in \mathbb{R}$.
This is a first-order linear ordinary differential
equation for the wave function $\Psi(k)$,
\begin{equation}
k^2 \Psi'(k)
+\left(k\left(1+{w}\right)-\bar{k}w-i \mu\right)\Psi(k) =0,
\end{equation}
where we have defined the dimensionless parameter
$\mu:=\frac{2 G M_0}{\hbar}\bar{M}$ to encode the expectation value of the mass $\bar M$.
The integration of this equation is straightforward.
However, we note that $k=0$ is a singular point,
and thus the equation does not couple
the regions $k>0$ and $k<0$. Therefore, the general solution reads,
\begin{align}
\Psi(k)=
\begin{cases}
A_-\,(-k)^{-(1+w)}\,e^{-(\bar{k}w+i \mu)/k}, & k<0,\\
A_+\,k^{-(1+w)}\,e^{-(\bar{k}w+i\mu )/k}, & k>0,
\end{cases}
\end{align}
with two independent integration
constants $A_-$ and $A_+$. Depending on the sign of
the product $\bar kw_r$, the exponential function is divergent
as $k$ approaches zero either from below or
from above. Therefore, $\Psi(k)$ to be normalizable,
one needs to fix either $A_-$ or $A_+$ to zero,
such that the state only has support either for $k>0$ or $k<0$.
In addition, the convergence of the integral involved in the computation of
$\langle \widehat{\cal O}_{p_u}^n\rangle$
requires $w_r>(n-1)/2$ and a fixed sign of
the product $\bar kw_r$. 

In conclusion, making use of the Heaviside function
$\Theta$, the most general intelligent state, which obeys conditions $i/$ and $ii/$ above,
can be written in a compact form as
\begin{equation}\label{eq.peakedstate}
\Psi(k)=\Theta(k\bar k)\frac{(2 w_r)^{w_r}}{\sqrt{\Gamma(2w_r)}}
|\bar{k}|^{w_r+\frac{1}{2}}\,
|k|^{-(1+w)}\,e^{-(\bar{k}w+i \mu)/k},
\end{equation}
with $w_r>1/2$ and $w_i\in\mathbb{R}$. The global coefficient has been chosen such
that the probability distribution defined by this wave function, $|\Psi(k)|^2$,
is normalized to one. Note that the sign of $\bar k$
is free, but it fixes whether the function has support for
positive or negative values of $k$.
In addition, the sign of the mass parameter $\mu$ is also free.
Projecting this state in the mass eigenstates \eqref{eq.masseigenstates},
yields its corresponding mass spectrum,
\begin{equation}
    \tilde{\Psi}(M)=\text{sgn}(\bar{k})\sqrt{\frac{GM_0}{\pi\hbar}}\frac{(2 w_r)^{w_r}}{\sqrt{\Gamma(2w_r)}}\Gamma(1+w)
|\bar{k}|^{w_r+\frac{1}{2}}\left(\bar{k}w-\frac{2iGM_0(M-\bar{M})}{\hbar}\right)^{-(1+w)}.
\end{equation}

\begin{figure}
\centering
\begin{subfigure}{0.49\textwidth}
    \includegraphics[width=\textwidth]{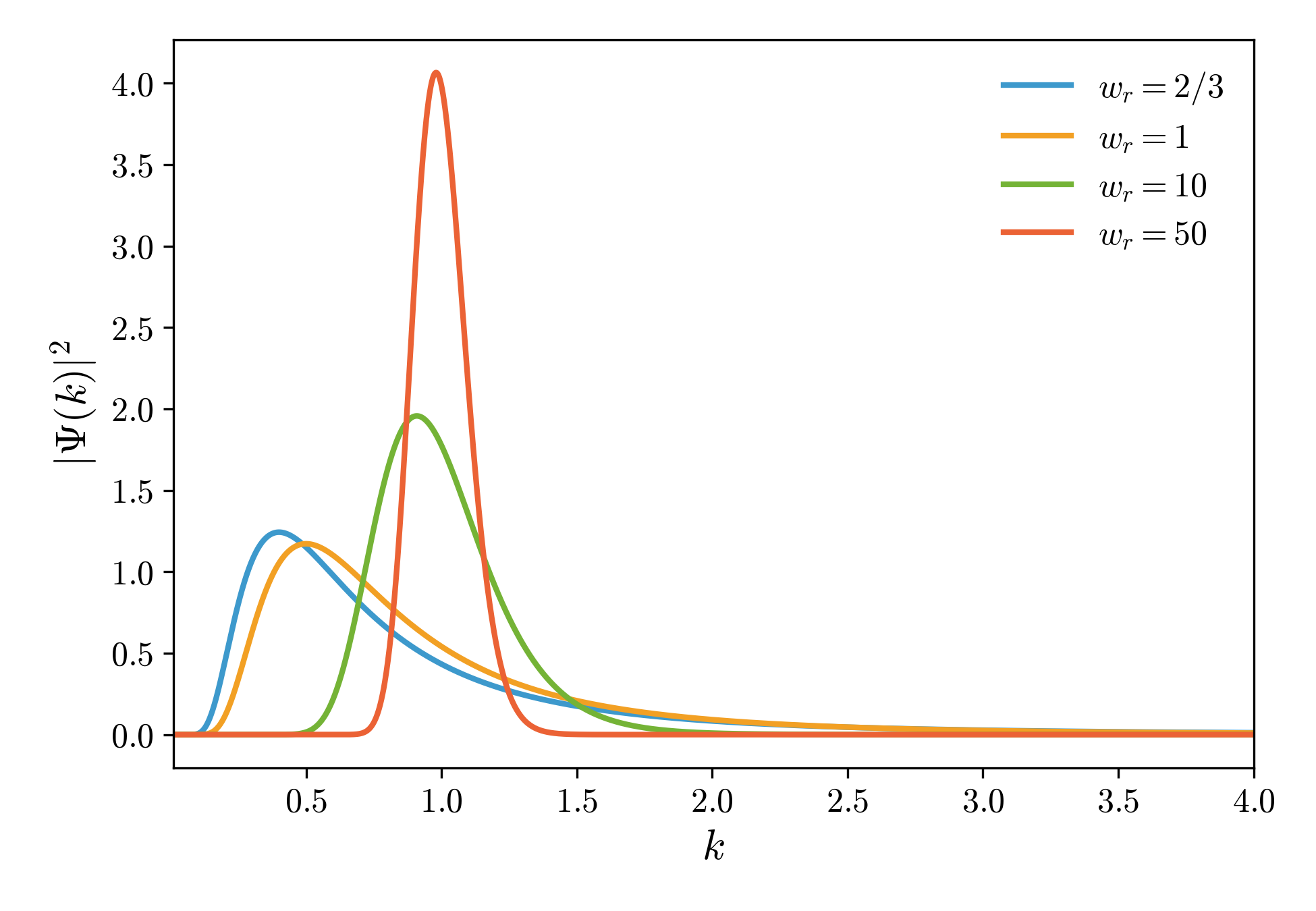}
    \end{subfigure}
\hfill
\begin{subfigure}{0.49\textwidth}
    \includegraphics[width=\textwidth]{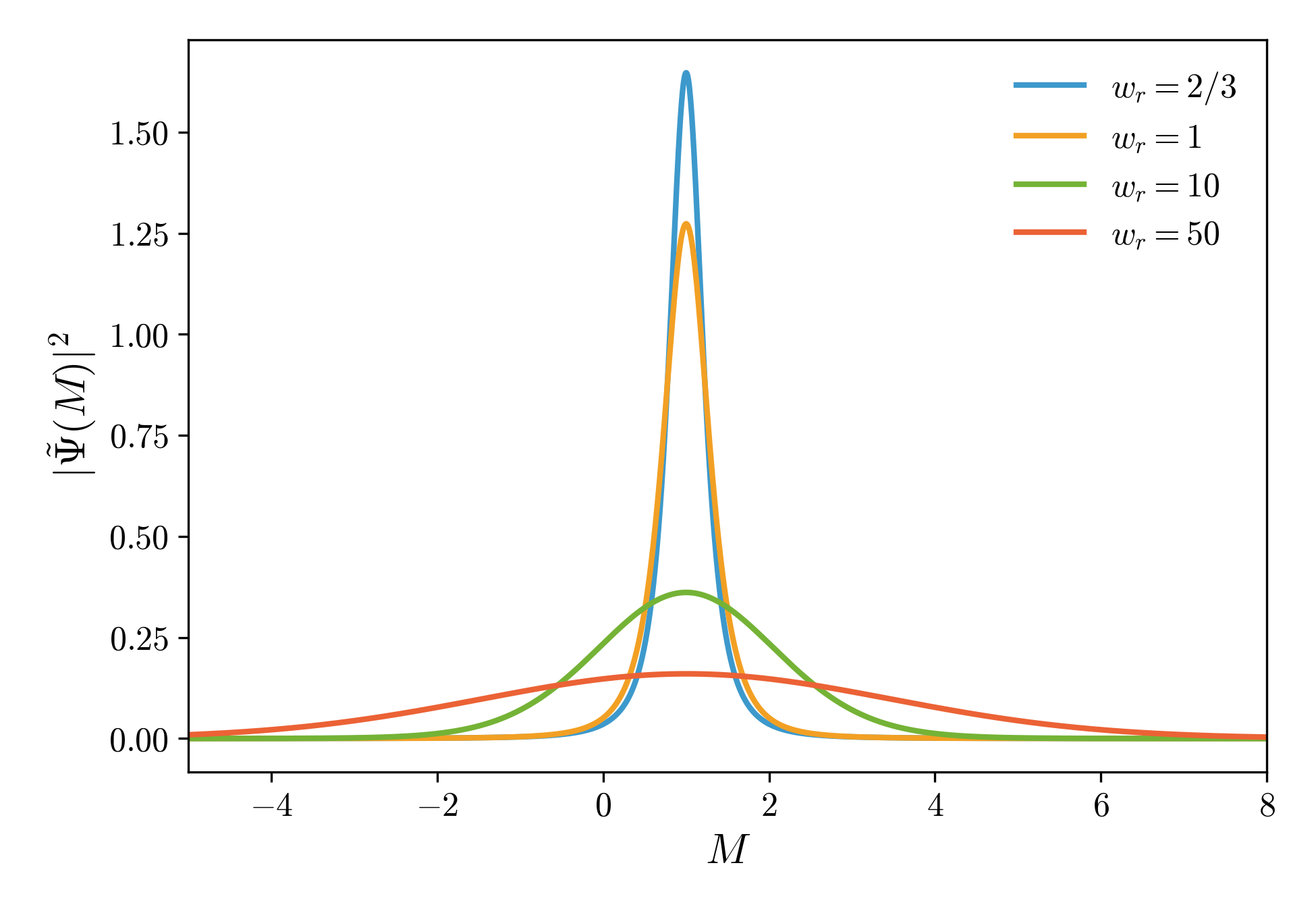}
\end{subfigure}
\caption{Probability distribution in the  $k-$space (left) and in the mass representation (right) for the intelligent state with parameters $\bar{k}=1$, $\bar{M}=1$, $w_i=0$, and different values of $w_r$.
Increasing $w_r$ makes the distribution in $k$ narrower, sharply localizing the state around the expectation value $\bar{k}=1$. Simultaneously, the distribution in $M$ becomes broader. Decreasing $w_r$ produces the opposite effect: the distribution in $k$ widens, whereas the distribution in $M$ becomes narrower and more sharply peaked around the expectation value $\bar{M}=1$.
\label{fig:figures2}}
\end{figure}
\begin{figure}
\centering
\begin{subfigure}{0.49\textwidth}
    \includegraphics[width=\textwidth]{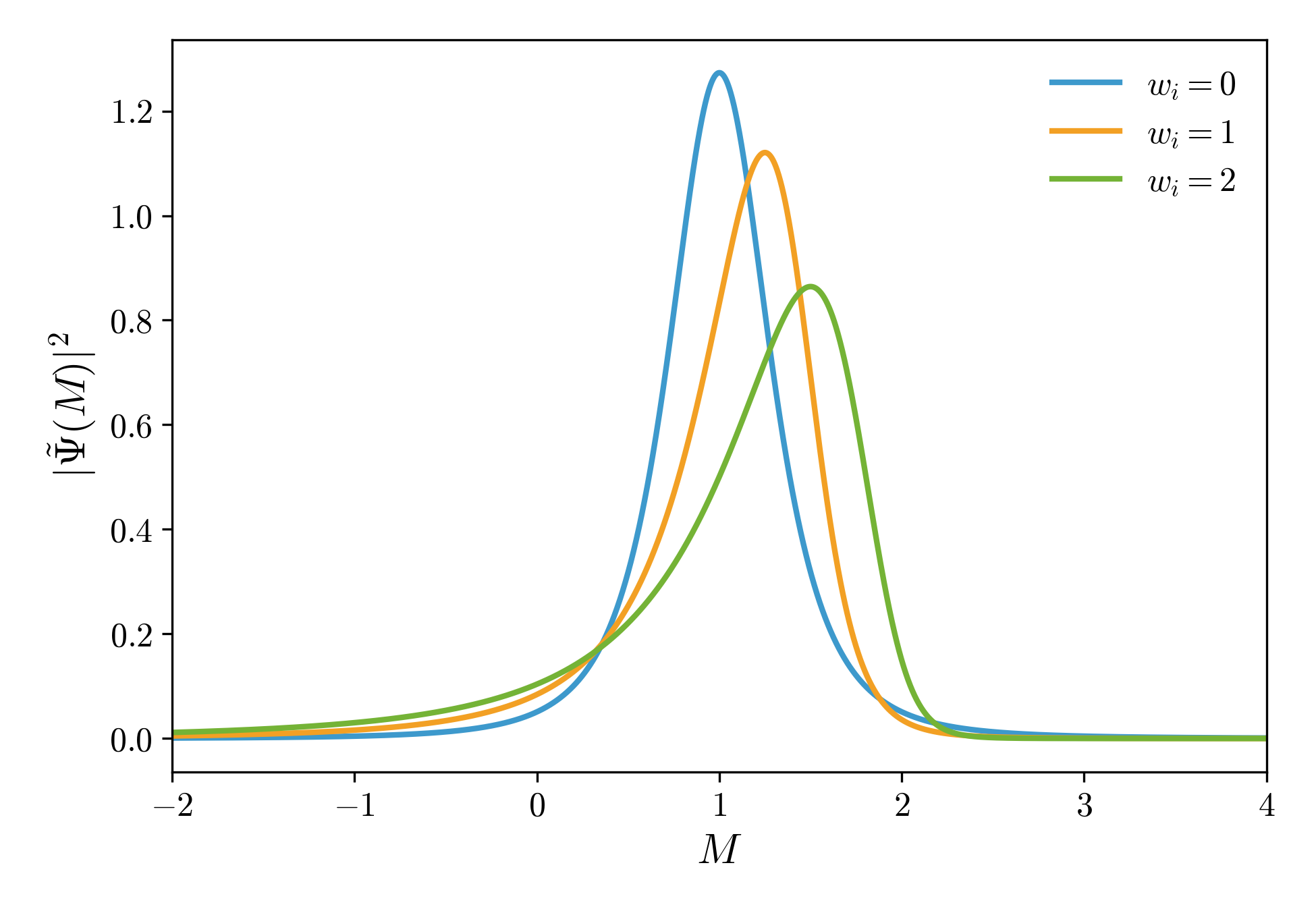}
\end{subfigure}
\hfill
\begin{subfigure}{0.49\textwidth}
    \includegraphics[width=\textwidth]{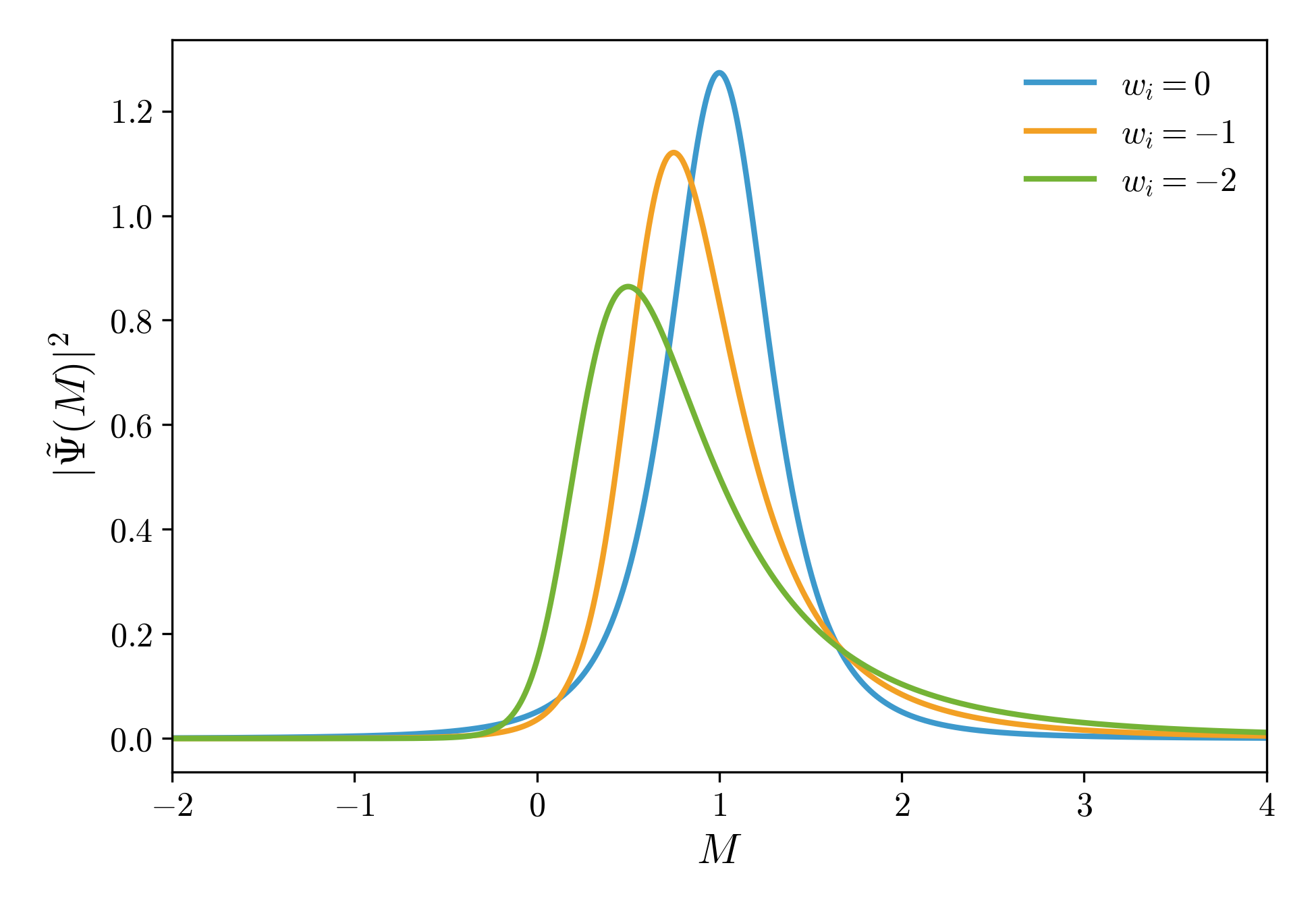}
\end{subfigure}
\caption{Probability distribution in the mass representation for the intelligent state with
parameters $\bar{k}=1$, $\bar{M}=1$, $w_r=1$, and different positive (left) and negative (right)
values of $w_i$. A positive (negative) $w_i$ shifts the peak of the distribution towards larger (smaller) values of
$M$, introducing an asymmetric tail.
\label{fig:figures3}
}
\end{figure}

In fact, for given $\bar M$ and $\bar k$, \eqref{eq.peakedstate}
does not define just an intelligent state, but a family of intelligent states parameterized
by a complex parameter $w$, or equivalently by its real and imaginary parts, $w_r$ and $w_i$.
These parameters determine the width and, more generally, the shape of the probability distributions around $\bar M$ and $\bar k$.
This can be seen in Figs.~\ref{fig:figures2}--\ref{fig:figures3}, where we plot the probability distribution $|\Psi(k)|^2$ in the $k$-space, as well as in the mass representation $|\tilde \Psi(M)|^2$, for different values of $w_r$ and $w_i$.
In particular, Fig.~\ref{fig:figures2} shows that, for a fixed $w_i$, increasing $w_r$ sharpens $|\Psi(k)|^2$ around the expectation value $\bar{k}$, while broadening $|\tilde\Psi(M)|^2$ around $\bar{M}$. On the other hand, as illustrated in Fig.~\ref{fig:figures3}, for a fixed $w_r$, a nonvanishing $w_i$ shifts the peak of $|\tilde\Psi(M)|^2$ away from $\bar{M}$ towards larger (smaller) values of $M$ for positive
(negative) $w_i$, and introduces an asymmetric tail. However, we note that
$|\Psi(k)|^2$ does not depend $w_i$, so its shape is invariant under changes of $w_i$.

In order to quantify these features and identify semiclassical states within the family \eqref{eq.peakedstate},
it is convenient to compute the second-order moments. Specifically, the relative fluctuations read
\begin{align}
\frac{\Delta\mathcal{O}_{p_u}}{|\langle\widehat{\mathcal{O}}_{p_u} \rangle|} &= \frac{1}{\sqrt{2w_r-1}},\label{eq:pufluct}\\\label{eq:massfluct}
\frac{\Delta\mathcal{O}_{M}}{|\langle\widehat{\mathcal{O}}_{M} \rangle|} &=\frac{|\bar k|}{|\mu|}\frac{|w|}{\sqrt{2w_r-1}},
\end{align}
where $|w|=\sqrt{w_r^2+w_i^2}$ is the modulus of $w$,
while the correlation is given by
\begin{equation}\label{eq.covariance}
\frac{1}{\Delta\mathcal{O}_{p_u}\Delta\mathcal{O}_{M}}\langle (\widehat{\cal O}_M-\langle\widehat{\cal O}_M \rangle)(\widehat{\cal O}_{p_u}-\langle\widehat{\cal O}_{p_u} \rangle)|_{\rm symm.} \rangle
=-\frac{w_i}{|w|}.
\end{equation}
At this point, we define the class of semiclassical intelligent states as those with small relative fluctuations \eqref{eq:pufluct}--\eqref{eq:massfluct},
which implies $1\ll w_r$ and $|\bar k \,w|\ll \sqrt{w_r}|\mu|$. Therefore,
for fixed $\bar k$, $w_r$, and $w_i$, semiclassicality requires a sufficiently large dimensionless mass $|\mu|$.
(Note that, at this level, the sign of $\mu$ is not fixed, and it could either be positive or negative.)
In particular, increasing $w_r$ or $|w_i|$ requires a correspondingly larger value of $|\mu|$ in order to maintain
small relative fluctuations. Finally, note also that the semiclassical states need not be uncorrelated:
according to Eq.~\eqref{eq.covariance}, they can exhibit a nonvanishing correlation determined by $w_i$.

\section{
Semiclassical description of the Schwarzschild black hole
}\label{sec.physicsintelligentstates}

We now employ the semiclassical intelligent states introduced above
to construct a semiclassical description of the Schwarzschild black hole
and analyze its physical implications.
This section is composed by two subsections. In Sec.~\ref{sec.quantumkilling} we evaluate
the expectation values of the different observables for the semiclassical intelligent states,
which, neglecting fluctuations,
allows us to define the location of the quantum Killing horizon and its basic geometric properties.
In Sec.~\ref{sec.thermodynamics} we study the thermodynamic properties of the quantum
horizon and describe its evaporation. In particular, we identify a natural state
that may describe a stable remnant as the end stage of the evaporation process.

\subsection{Quantum Killing horizon and curvature invariants}\label{sec.quantumkilling}

In a semiclassical regime, the expectation value of the different observables
can be interpreted to provide small modifications to the classical geometry.
In this way, semiclassical intelligent states can provide us with a semiclassical
description of a quantum black hole $(\mu>0)$ or a naked Schwarzschild singularity $(\mu<0)$.
The flat Minkowski geometry $(\mu=0)$ is however excluded from this semiclassical description,
as its corresponding relative mass fluctuations \eqref{eq:massfluct} would be formally infinite.
Since we are particularly interested in analyzing quantum effects on the black-hole horizon,
in the following we will assume $\mu>0$, and thus $\bar M>0$.

Let us thus begin by analyzing the expectation values of the different observables
for states \eqref{eq.peakedstate} with
$1\ll w_r$, $|\bar k \,w|\ll \sqrt{w_r}\mu$, and $\mu>0$.
By construction, the expectation values of the relational observables
${\widehat{\cal O}}_M$ and ${\widehat{\cal O}}_{p_u}$ are $\bar M$ and $\bar k M_0$, respectively.
Concerning the observables related to the curvature scalars, from \eqref{eq:relRopp}, we recall
that for any physical state $\langle {\widehat{\cal O}}_R\rangle=0$, while, from \eqref{eq:relPSopp},
we find,
\begin{equation}\label{eq.expectationvaluepsi2}
\langle\widehat{\mathcal{O}}_{\Psi_2}\rangle =-\frac{G\m0}{\ell^3 e^{3\sigma}}=-\frac{G\m0}{\langle\hat{\cal O}_r\rangle^3}.
\end{equation}
Therefore, at the level of expectation values (disregarding fluctuations), this quantum state exactly reproduces
the curvature of the Schwarzschild black hole with mass parameter $\bar M$, including the
curvature divergence as $\langle\widehat{\cal O}_r\rangle\to 0$.

In order to analyze the causal structure of this semiclassical description of the Schwarzschild black hole,
we consider now the expectation value of the squared norm of the Killing \eqref{eq:relkillopp},
which, for this family of states, reads
\begin{equation}
\langle\widehat{\mathcal{O}}_\xi\rangle=\frac{(1+w_r)(1+2w_r)}{2\bar{k}^2w_r^2}\left(-1+\frac{1}{\ell e^\sigma}\left(2G\bar{M}+\frac{\hbar w_i \bar{k}}{M_0(1+w_r)}\right)\right),
\end{equation}
and it depends on all the parameters of the state.
At this point, we recall that at the classical level $k$ is a Dirac observable related to coordinate choices,
and, more precisely, to the lapse of the static observers at infinity. Hence, in order to fix 
$\bar k$ in a meaningful way at this semiclassical level, we request the Killing field to be normalized
to unity at asymptotic infinity,
that is, $\langle\widehat{\mathcal{O}}_\xi\rangle\to -1$ as $\langle\widehat{\mathcal{O}}_r\rangle\to\infty$.
This condition fixes the norm of $\bar k$ as $\bar k^2={(1+w_r)(1+2w_r)}/{(2w_r^2)}$,
though, as in the classical model,
the sign of $\bar k$ is still free. With this normalization,
the semiclassicality conditions can be written in a simpler form as the requirements $1\ll w_r$ and
$1\ll\frac{|w|^2}{w_r}\ll \mu^2$, which implies relatively large values of $w_r$ and $\mu$,
while $w_i$ can not take extremely large values in module, i.e., $|w_i|\ll\sqrt{w_r}\mu\ll\mu^2$.

In this way,
the squared norm of the Killing can now be expressed in the particularly illuminating form,
\begin{equation}
    \label{eq:sqnk}
    \langle\widehat{\mathcal{O}}_\xi\rangle=-1+\frac{r_h}{\ell e^\sigma},
\end{equation}
where we have defined
\begin{equation}\label{def.meff}
    r_h:=
2 G\bar M
\left(1+
\text{sgn}(\bar k)\,
\frac{w_i}{2 \mu}
\frac{\sqrt{2(1+w_r)(1+2w_r)}}{w_r(1+w_r)}\right),
\end{equation}
which, considering that $1\ll w_r$, can be approximated as
\begin{equation}
r_h\approx2 G\bar M \left(1+\text{sgn}(\bar k)\frac{w_i}{\mu w_r}\right).
\end{equation}
Recalling that $\langle \widehat {\cal O}_r \rangle=\ell e^\sigma$, from \eqref{eq:sqnk}
it is straightforward to see
that the expectation value of the Killing $\langle\widehat{\mathcal{O}}_\xi\rangle$ vanishes
when $\langle \widehat {\cal O}_r \rangle=r_h$. Therefore, we interpret $r_h$ as the radius
of the quantum (semiclassical) Killing horizon.
In this way, if the state under consideration is uncorrelated $(w_i=0)$, we observe
that $r_h$ reproduces the classical Schwarzschild radius $2G \bar M$ for a black hole of mass $\bar M$.
However, the horizon radius defined by a correlated state, with a nonvanishing $w_i$,
is shifted with respect to its classical value towards smaller (for $\bar k w_i<0$)
or larger (for $\bar k w_i>0$) values.

Concerning the quantities that characterize the gravitational pull, using \eqref{eq.acceleration},
and considering the value of $|\bar k|$ as fixed above,
we can compute the effective proper acceleration experienced by a static observer,
\begin{equation}\label{eq.aeff}
a_\text{eff}=a_{0}(r_h)\left(1+\frac{\hbar^2}{M_0^2 r_h^2}\frac{({w_i^2+(1+w_r)(5+w_r)})}{2({1+w_r})}\right)^{1/2},
\end{equation}
where $a_0(R):=R/(2\ell^2e^{2\sigma}\left(1-\frac{R}{\ell e^\sigma}\right)^{1/2})$
is the classical proper acceleration for a observer at a radius $R$.
Since the second term inside the brackets is positive definite,
for a given black hole with a fixed horizon radius $r_h$, quantum effects increase
the hovering acceleration as compared to its classical value.
This implies that quantum degrees of freedom act as positive energy modes of the black
hole and contribute to its gravitational pull.
Considering the order of magnitude of the parameters for the semiclassical state, this expression
can be approximated by
\begin{equation}
a_\text{eff}
\approx
a_{0}(2 G \bar M)\left(1+\frac{5}{4\mu^2}
+{\rm sgn}(\bar k)\frac{w_i}{w_r\mu}\frac{(\ell e^\sigma-G \bar M)}{(\ell e^\sigma-2 G \bar M)}\right)
.
\end{equation}
This expression is more suited to compare the proper acceleration for a static observer
required by a classical and a quantum black hole with the same mass parameter $\bar M$. In this case,
the quantum modifications also produce an increase of the acceleration, although correlated states
with $\bar k w_i<0$ may produce a decrease of the acceleration depending on the specific values
of the different parameters.

We can do a similar computation for the surface gravity, that is, the definition
\eqref{eq.kappa} leads to 
\begin{align}\label{eq.kappaeff}
\kappa_\text{eff}=\kappa_{0}(r_h)\left(1+\frac{\hbar^2}{M_0^2 r_h^2}\frac{(w_i^2+(1+w_r)(5+w_r))}{2(1+w_r)}\right)^{1/2}
\approx \kappa_{0}(2 G \bar M) \left(1+\frac{w_r}{4 \mu^2}-\,{\rm sgn}(\bar k)\frac{w_i}{\mu w_r} \right),
\end{align}
with $\kappa_{0}(R):=1/(2 R)$ being the classical surface gravity of a black hole
with horizon radius $R$. As with the hovering acceleration, quantum effects contribute
to enhance the surface gravity, as compared with a classical black hole with the same
horizon size $r_h$. If we compare a quantum and a classical black hole with the
same mass parameter $\bar M$, in general the surface gravity is also increased by quantum
effects, except for certain correlated states with $(\bar k w_i)>0$ and $w_r^2<4\mu|w_i|$.

\subsection{Thermodynamic properties and evaporation of the quantum horizon}
\label{sec.thermodynamics}

A fundamental result of quantum field theory in curved spacetimes is that black holes behave
as genuine thermodynamic systems. Hawking's semiclassical analysis showed that quantum vacuum fluctuations in the presence of
the event horizon of a black hole give rise to the emission of thermal radiation,
implying that one can associate a temperature to a black hole, known as the Hawking temperature \cite{Hawking}.
For stationary spacetimes admitting a regular Killing horizon, this result acquires a universal geometric interpretation:
the Hawking temperature is completely determined by the classical value of the surface gravity $\kappa$
according to
\begin{equation}\label{sec.classicalT}
T=\frac{\hbar\kappa}{2\pi},
\end{equation}
where we have set the Boltzmann constant to $k_B=1$. This relation is independent of the particular black-hole geometry,
and therefore applies to any stationary geometry possessing a regular Killing horizon \cite{GibbonsPerry1978}.

In the derivation of this result, a classical horizon is assumed. Thus, one
may replace $\kappa$ by its form in terms of the classical horizon radius or
in terms of the black-hole mass, and the value of the temperature would be unaltered.
However, in the semiclassical picture developed in the previous subsection, it is clear that the classical relations among
the expectation value of the mass operator $\bar M$, the (effective) surface gravity $\kappa_{\rm eff}$,
and the radius of the Killing horizon $r_h$, as defined by the vanishing of the expectation value of the Killing field,
need not hold. That is, in general
we have that $\kappa_{\rm eff}\neq 1/(2 r_h)$ and $r_h\neq 2 \pi G \bar M$.
Therefore, if the temperature is defined through any one of these three quantities,
adjusting the expression \eqref{sec.classicalT} making use of classical relations,
one obtains, in general, different results.
More specifically, one may define,
\begin{align}\label{eq.Tkappa}
T_{\kappa} &:=\frac{\hbar\, \kappa_{\rm eff}}{2\pi}
=\frac{\hbar}{4\pi r_h}\sqrt{1+\frac{\hbar^2}{M_0^2 r_h^2}\frac{(w_i^2+(1+w_r)(5+w_r))}{2(1+w_r)}}
\approx\frac{\hbar}{8\pi G \bar M} \left(1+\frac{w_r}{4 \mu^2}-\,{\rm sgn}(\bar k)\frac{w_i}{\mu w_r} \right),\\\label{eq.Trh}
T_{r_h} &:=\frac{\hbar}{4 \pi r_h}\approx\frac{\hbar}{8\pi G \bar M}\left(1-\,{\rm sgn}(\bar k)\frac{w_i}{\mu w_r} \right),\\\label{eq.TM}
T_{\bar M} &:=\frac{\hbar}{8\pi G \bar M}.
\end{align}
As expected, these three definitions coincide in the limit $\mu\to\infty$, which corresponds to black holes of infinite mass and
thus vanishing relative mass fluctuations \eqref{eq:massfluct}, but generally differ for finite $\mu$.

At this level, it is therefore not possible to determine unambiguously which, if any, of these
quantities should be identified with the thermodynamic temperature of the semiclassical horizon.
Establishing such an identification would require coupling the semiclassical horizon to an
additional quantum field and analyzing its thermodynamic properties. Such an analysis
lies beyond the scope of the present simplified model.
Therefore, for definiteness, in what follows we postulate the temperature of the horizon to be $T_\kappa$.
This choice is motivated by the fact that, as commented above, the Hawking temperature
is fundamentally related to the surface gravity through \eqref{sec.classicalT},
whereas the expressions involving the horizon radius \eqref{eq.Trh} or the mass \eqref{eq.TM} rely on the classical Schwarzschild relations between these quantities and the surface gravity.

In addition to the temperature,
in order to derive the entropy of the quantum black hole, we also need to specify the energy contained inside the Killing horizon. We naturally identify this energy with the expectation value of the mass observable $\widehat{\mathcal{O}}_M$, namely, $E:=\langle \widehat{\mathcal{O}}_M \rangle=\bar{M}$.
In this way, we can directly use the definition of entropy $\delta S=\delta E/T_\kappa$, and write
\begin{equation}
    S=S_0+\int \frac{\D \bar{M}}{T_\kappa(\bar{M})}.
\end{equation}
Making use of the relation \eqref{def.meff} between $\bar M$ and the horizon radius $r_h$,
it is straightforward to express this as an integral over $r_h$,
\begin{equation}
    S=S_0+\frac{1}{2G} \int\frac{\D r_h}{T_\kappa(r_h)},
\end{equation}
where one can use the temperature as given in \eqref{eq.Tkappa}.
Performing a series expansion for large horizon areas $A:=4\pi r_h^2$, and fixing the integration constant $S_0$ appropriately, we find
\begin{equation}\label{eq.entropy}
S=\frac{A}{4G\hbar}-\frac{\hbar\pi(w_i^2+(1+w_r)(5+w_r))}{4GM_0^2(1+w_r)}\,\ln A+{\cal O}\left(\frac{\hbar^3}{M_0^4 G A}\right).
\end{equation}
It is interesting to note that the dominant contribution coincides with the Bekenstein-Hawking entropy, whereas the leading quantum correction
is logarithmic in the horizon area. Logarithmic corrections of this form arise generically in a broad class of approaches to quantum gravity \cite{Kaul,Meissner:2004ju,Sen:2011ba,Sen:2012dw}.

The description of the quantum black hole we have provided so far
can be considered accurate in the semiclassical regime, where relative quantum fluctuations
are negligible, which particularly implies a large dimensionless mass $\mu$.
However, as the black hole evaporates, its mass decreases and, once $\mu$
can no longer be considered to be
very large, the contribution of the fluctuations must be taken into account.
In particular,
in this approach to a deeper quantum regime, it is natural to identify the horizon energy with the expectation value of the mass
together with its quantum fluctuation, that is, $E:=\bar{M}+\Delta {\cal O}_M$. As can be seen in Eq.~\eqref{eq:massfluct}, this
fluctuation depends on the parameters $w_r$ and $w_i$ that characterize the quantum state, and, interestingly,
it attains a minimum for the specific values $w_r=1$ and $w_i=0$.
Assuming that the emission of Hawking radiation drives the system towards the configuration of minimal energy, it is therefore natural to
identify the state with $\bar{M}=0$, $w_r=1$, and $w_i=0$ as the final remnant state of the evaporation process.

Concerning the quantum geometry of this remnant state, given these set of parameters, it is easy to see that its corresponding $r_h$, as defined in \eqref{def.meff}, is zero, and thus, the expectation value of the squared Killing norm \eqref{eq:sqnk} never vanishes.
In this sense, one would conclude that this is a horizonless object.
However, as already commented, this remnant state lies in a regime where quantum fluctuations can no longer be neglected,
and the quantum geometry, defined in terms of expectation values, ceases to provide a complete characterization of the system.
Since the fluctuations are large, they induce an uncertainty in the different observables
and, in particular, in the norm of the Killing field.
Thus one can not interpret the horizon to be defined
on a specific value of the radius. At most the notion of horizon should be generalized to a shell
with a finite width determined by the fluctuation
\begin{equation}
\Delta r_h:= \frac{\D\langle\widehat{\mathcal{O}}_r\rangle}{\D\langle\widehat{\mathcal{O}}_\xi\rangle}\Delta\mathcal{O}_\xi\,\bigg|_{\sigma=\sigma_h},
\end{equation}
which is defined by standard error propagation, with $\Delta\mathcal{O}_\xi$ being the fluctuation of
$\widehat{\mathcal{O}}_\xi$ and $\sigma_h$ being defined by the equation
$\langle \widehat{\mathcal{O}}_\xi \rangle\big|_{\sigma=\sigma_h}=0$, or, equivalently, by $r_h=\ell e^{\sigma_h}$.
Specifically, for a general intelligent state with parameters $(\bar M, w_r, w_i)$, one has the explicit expression,
\begin{equation}
\Delta r_h=\frac{\hbar}{M_0}\frac{\sqrt{2(2+w_r)((1+w_r)^2+w_i^2)}}{2(1+w_r)},
\end{equation}
and thus its corresponding ``diffused horizon''
is located in the radial interval $(r_h-\Delta r_h, r_h+\Delta r_h)$,
with $r_h$ as given in \eqref{def.meff}.
Therefore, for the specific parameters of the remnant state,
$r_h= 0$, $w_r=1$, and $w_i=0$, the diffused horizon extends
through the radial interval $(0,\sqrt{\frac{3}{2}}\frac{\hbar}{M_0})$.
In summary, the remnant state is interpreted as an object with a diffused horizon
of a size $\sqrt{\frac{3}{2}}\frac{\hbar}{M_0}$ and energy
$\bar M+\Delta M=\frac{\sqrt{3}}{2}\frac{\hbar}{GM_0}$.

\section{Conclusions}\label{sec:Conclusions}

We have considered the canonical quantization of the Schwarzschild geometry in a minisuperspace approach,
where a radial foliation of the Schwarzschild spacetime is considered. Because of staticity and spherical symmetry,
the resulting model features degrees of freedom that depend only on the radial variable. This description allows
us to treat both the interior and the exterior regions of the black hole within a common framework, and, in particular,
investigate physical effects in the vicinity of the horizon.
The main results of this paper consist, on the one hand, of a precise construction of the physical Hilbert space,
together with a set of gauge-invariant relational observables, and, on the other hand, of a semiclassical description of the
Schwarzschild black hole in terms of the so-called intelligent states, including the description of its thermodynamic properties
and evaporation process.

More precisely, concerning the mathematical description, although different works in the literature have considered the quantization of this model
before (see, e.g., Refs.~\cite{Kuchar:1994zk,Kastrup:1993br,ThiemannKastrup:1993,Brotz:97,Kenmoku:1999,Kenmoku:1999bf,Cavaglia:1994yc, Cavaglia:1995bb}),
an outstanding issue is the precise construction of the Hilbert space, its inner product, and the calculation of expectation values of suitable observables.
The main difficulty is to construct these objects directly from the solutions to the quantum constraint equation, without an a priori choice of gauge, thus respecting
the local symmetries of the model.
In this paper, we put forward a straightforward approach to the quantization of this model following previous developments in the literature of relational observables and
physical inner product \cite{Chataignier:2019kof,Chataignier:2020fys,Chataig:Thesis}. We discuss how the canonical quantization of this model
can be understood in terms of a physical Hilbert space, which is the space of solutions to the quantum constraint equation, endowed with the appropriate inner product, obtained by the Rieffel induction procedure \cite{Halliwell:91,Ashtekar:94,Landsman:95,Marolf:97,Embacher:98,Giulini:99,Giulini:99-2,Giulini:2000,Marolf:2000,Halliwell,Chataignier:2019kof,Chataignier:2020fys,Chataig:Thesis}.
In addition, we construct quantum relational observables describing the geometry, mass content, and gravitational pull of the quantum black hole.
These observables are gauge invariant, in the sense that they commute with the constraint operator, and they encode the value of physical quantities with respect to a chosen radial variable.

Having a Hilbert space of physical states and appropriate invariant observables, one can then proceed to compute expectation values and correlation functions.
In order to construct a natural set of states suitable to provide a semiclassical description of the Schwarzschild black hole,
we introduce the so-called intelligent states, which saturate the Robertson--Schr\"odinger uncertainty relation
for the observables related to the mass and the asymptotic Killing norm. This construction yields the family of intelligent states
\eqref{eq.peakedstate}, characterized by four free parameters: two specifying the expectation value of the mass $(\bar M)$ and of the asymptotic Killing norm $(\bar k)$,
and the other two $(w_r$ and $w_i$) encoding the fluctuations and correlation of the state. Within this family, we define the semiclassical
states as those with small relative fluctuation \eqref{eq:pufluct}--\eqref{eq:massfluct}, which imposes a certain hierarchy between
the values of the different free parameters.

We then present a description of a semiclassical Schwarzschild black hole in terms of such states, and obtain the leading
quantum corrections to the classical model by computing expectation values of the relational
observables associated with different quantities. In particular, the expectation values
of the observables associated with the curvature invariants
exactly reproduce their classical expressions: a vanishing Ricci curvature and a Newman--Penrose $\Psi_2$ scalar \eqref{eq.expectationvaluepsi2}
proportional to $1/r^3$. However, concerning the radial location of the black-hole horizon \eqref{def.meff},
quantum corrections introduce a shift proportional to the correlation parameter $w_i$.
This shift can be either positive or negative, and it only vanishes for uncorrelated states with $w_i=0$.
We also compute the effective hovering acceleration of a static observer outside the semiclassical black hole \eqref{eq.aeff},
as well as its surface gravity \eqref{eq.kappaeff}. It turns out that, as compared with a classical black hole with the same
horizon size $r_h$, both these quantities are enhanced by the quantum effects. Therefore, in this sense,
quantum degrees of freedom contribute in a positive way to the gravitational pull of the black hole.

Concerning the thermodynamic properties of the semiclassical black hole,
in this simple model it is unclear how to define the temperature
and the energy content of the horizon, as different definitions, which are equivalent
making use of classical identities, lead to different expressions in the quantum setting.
For definiteness, we postulate that temperature is given in terms of the surface gravity \eqref{eq.Tkappa},
while the energy content is encoded in the expectation value of the mass operator.
In this way, we compute the entropy of the semiclassical black hole \eqref{eq.entropy}, which,
at leading order, reproduces the well-known Bekenstein--Hawking expression.
It is interesting to note that the dominant quantum correction to the entropy
is proportional to the logarithm of the event horizon area, a feature that also appears
in many other approaches to quantum gravity \cite{Kaul,Meissner:2004ju,Sen:2011ba,Sen:2012dw}.
Finally, we analyze the evaporation process associated with Hawking radiation and identify a
state of minimal energy with a nonvanishing horizon radius.
This state may be interpreted as a stable quantum remnant left behind at the end of the evaporation process.

The model discussed here can serve as a starting point for more realistic models of quantum black holes.
In particular, the different technical developments, such as the construction of a physical Hilbert
space with appropriate gauge-invariant observables and the definition of natural states
providing a semiclassical description of a black hole, can be useful in more realistic settings.
Also, the thermodynamic analysis presented here points to the possibility that the canonical formalism
may provide a consistent and rigorous framework to address the thermodynamics
of black holes and the related information problem within a suitable class of states.

\section*{Acknowledgments}

DB and MLE acknowledge financial support by the Basque Government
Grant \mbox{IT1977-26}, and by the Grant PID2021-123226NB-I00 (funded by
MCIN/AEI/10.13039/501100011033 and by ``ERDF A way of making Europe'').

\appendix

\section{\label{app:null}Null expansions}

In this appendix we explicitly compute the null expansions of the normal
vectors to the spheres of constant $x$ and $t$ in the spacetime endowed with
the metric \eqref{eq:le}. For such construction, one begins by
arbitrarily introducing a positive direction of time. Since we are interested in charts
that are well defined at the horizon, a necessary condition is that
$N^2-s^2=c(1-2GM/r)+{\cal O}((1-2GM/r)^2)$ in a neighborhood of
$r=2GM$ for certain constant $c$. In particular, this additionaly implies that $s$
can not vanish at the horizon since $s=0$ would imply $N=0$
and thus a degenerate metric. Therefore, let us assume $s\neq 0$ in the following.

In such case, in order to fix the positive direction of time,
we can simply choose the timelike geodesic vector at rest at infinity,
\begin{equation}
v=|k|\frac{1-\frac{|s|}{N}\sqrt{\frac{2GM}{r}}}{\left(1-\frac{2GM}{r}\right)}\partial_t-
\frac{{\rm sgn}(s)}{N|k|}\sqrt{\frac{2GM}{r}}\,\partial_x,
\end{equation}
such that a causal vector $\omega$ will be future-pointing
if $g(v,\omega)<0$. This vector is well defined everywhere, and,
in particular, in the near-horizon regime, it takes the form,
\begin{equation}
 v\approx\frac{c+s^2}{2 s^2} |k|\partial_t-\frac{1}{|k|s}\partial_x.
\end{equation}
Depending the sign of $s$, this observer is in free fall towards increasing $(s<0)$
or decreasing $(s>0)$ values of $x$.

If $N^2-s^2=0$ for all values of $r$, the null vectors normal to the spheres can be defined as,
\begin{align}
n_1 &=-\frac{1}{s} \partial_x,\\
n_2 &=\partial_t+\frac{1}{2k^2s}\left(1-\frac{2GM}{r} \right) \partial_x,
\end{align}
such that they are both future pointing and normalized $g(n_1,n_2)=-1$.
The expansions of these vectors are defined as
\begin{equation}
 \theta_{(i)}:=\frac{1}{\sqrt{{\rm det}(q)}}\,n_i^a \nabla_a\left(\sqrt{{\rm det}(q)}\right),
\end{equation}
with $\sqrt{{\rm det}(q)}=r^2 \sin\theta$ being the volume element given by the determinant of the induced metric on the
spheres,
\begin{equation}
 \theta_{(i)}=\frac{2}{r}n_i^a \nabla_a r=\frac{2}{r}n^1_i r'(x).
\end{equation}
It is straightforward to check the signs of the expansions,
\begin{align}
 &{\rm sgn}(\theta_{(1)})=- {\rm sgn}(s\, k),\\ &{\rm sgn}(\theta_{(2)})={\rm sgn}(s\, k(r-2GM)),
\end{align}
such that ${\rm sgn}(\theta_{(1)}\theta_{(2)})=-{\rm sgn}(r-2GM)$. Hence, the region
$r>2GM$ is nontrapped, while $r<2GM$ is trapped (to the future) if $s\,k>0$ and
antitrapped (trapped to the past) if $s\,k<0$.

If $N^2-s^2=0$ only at $r=2GM$, a complete set of future-pointing
null vectors is given by
\begin{align}
n_1 &=-\frac{k^2}{2N^2}\left(\frac{N^2-s^2}{1-\frac{2GM}{r}}\right)\partial_t-\frac{1}{2N^2}(s+{\rm sgn}(s)N)\,
\partial_x,\\
n_2 &=\partial_t+\left(\frac{1-\frac{2GM}{r}}{N^2-s^2}\right)\frac{1}{k^2}(-s+{\rm sgn}(s)N)\,\partial_x,
\end{align}
such that the normalization $g(n_1,n_2)=-1$ holds. Note that both these vectors
are well defined at the horizon since the ratio $(N^2-s^2)/(1-\frac{2GM}{r})=c$ there.
 In the near horizon regime they take the form,
\eq{
n_1 &\approx\frac{c k^2}{2s^2}\partial_t-\frac{1}{s}\,\partial_x,\\
n_2 &\approx\partial_t+\frac{r-2M}{4Msk^2}\,\partial_x.
}

Defining the expansions as above,
we obtain a similar result concerning the trapeness of the different regions:
\eq{
 {\rm sgn}(\theta_{(1)})&=- {\rm sgn}(s\, k),\\
 {\rm sgn}(\theta_{(2)})&={\rm sgn}(s\, k(r-2GM)),
}
such that ${\rm sgn}(\theta_{(1)}\theta_{(2)})=-{\rm sgn}(r-2GM)$. Hence, the region
$r>2GM$ is nontrapped, while $r<2GM$ is trapped if $s\,k>0$ and
antitrapped if $s\,k<0$.

\section{Unitary representations of the dilation group}\label{app:dilationgroup}

We study the representation of the multiplicative group of positive numbers on  the Hilbert space
\[
L^2\left(\mathbb{R}^+\!, z^{2\gamma+1}\,dz\right),
\]
where $\gamma \in \mathbb{R}$ is a real parameter. For each $a \in (0, \infty)$, define the dilation operator $\widehat{D}_a$ acting on wave functions $\psi(z)$ by
\begin{equation}
    (\widehat{D}_a \psi)(z) := a^{\gamma+1}\, \psi(a z).
\end{equation}
This family of operators satisfies the group relations
\begin{equation}
    \widehat{D}_a \widehat{D}_{a'} = \widehat{D}_{a a'}, \qquad
    \widehat{D}_1 = \widehat{I}, \qquad
    \widehat{D}_a^{-1} = \widehat{D}_{a^{-1}},
\end{equation}
and preserves the inner product, since
\begin{equation}
    (\widehat{D}_a \psi, \widehat{D}_a \phi) = (\psi, \phi).
\end{equation}
Therefore, the map $a \mapsto \widehat{D}_a$ defines a unitary representation of the multiplicative group of positive real numbers on $L^2(\mathbb{R}^+, z^{2\gamma+1}dz)$.  
\\

The infinitesimal generator of this representation, is given as, usually, by
\begin{equation}
     -i \frac{d}{da}(\widehat{D}_a\psi)(z)\big|_{a=1} = -i \left(z\,\partial_z + \gamma + 1\right)\psi(z).
\end{equation}
By Stone's theorem, this operator is self-adjoint on a suitable dense domain.

\printbibliography

\end{document}